\documentclass[12pt]{article}
\usepackage[margin=1in]{geometry}
\usepackage[authoryear,round]{natbib}

\usepackage[T1]{fontenc}
\usepackage[utf8]{inputenc}
\usepackage{amsmath,amsthm,booktabs,graphicx,placeins,microtype,setspace}
\graphicspath{{figures/}}
\usepackage[colorlinks=true,citecolor=blue,linkcolor=blue,urlcolor=blue]{hyperref}
\usepackage{newtxtext}
\usepackage[subscriptcorrection,varvw]{newtxmath}
\theoremstyle{plain}
\newtheorem{theorem}{Theorem}
\newtheorem{proposition}{Proposition}

\newtheorem{corollary}{Corollary}
\newtheorem{assumption}{Assumption}
\theoremstyle{definition}
\newtheorem{example}{Example}
\newcommand{\E}{E}
\DeclareMathOperator{\pr}{pr}
\DeclareMathOperator{\var}{var}
\DeclareMathOperator{\cov}{cov}

\newcommand{\T}{\mathrm{\scriptscriptstyle T}}
\newcommand{\norm}[1]{\lVert #1\rVert}
\newcommand{\op}{o_{\mathrm p}}
\newcommand{\Op}{O_{\mathrm p}}
\newcommand{\R}{\mathbb R}
\newcommand{\Hh}{\mathcal H}
\newcommand{\Ff}{\mathcal F}

\newcommand{\Fc}{\mathcal C}
\newcommand{\Ii}{\mathcal I}
\newcommand{\ind}{\mathbf 1}

\newcommand{\external}[1]{\ifcsname external@#1\endcsname\csname external@#1\endcsname\else ??\fi}

\expandafter\def\csname external@app:sim-lasso\endcsname{S6.2}
\expandafter\def\csname external@app:sim-neural\endcsname{S6.3}
\expandafter\def\csname external@app:sim-power\endcsname{S6.5}
\expandafter\def\csname external@app:sim-replacement\endcsname{S6.7}
\expandafter\def\csname external@app:sim-training\endcsname{S6.4}
\expandafter\def\csname external@cor:projection\endcsname{S1}
\expandafter\def\csname external@ex:failure\endcsname{S1}
\expandafter\def\csname external@lem:haccompare\endcsname{S2}
\expandafter\def\csname external@lem:moments\endcsname{S1}
\expandafter\def\csname external@lem:replacement\endcsname{S3}
\expandafter\def\csname external@prop:lasso-approximate\endcsname{S2}
\expandafter\def\csname external@prop:mean-correction\endcsname{S1}
\expandafter\def\csname external@prop:nn-correction\endcsname{S3}
\expandafter\def\csname external@prop:nn-validation\endcsname{S4}
\expandafter\def\csname external@sec:degeneracy\endcsname{S5}
\expandafter\def\csname external@sec:nn-proofs\endcsname{S4.3}
\expandafter\def\csname external@sec:states\endcsname{S3.1}
\begin{document}
\title{Orthogonal inference for machine learning forecasts}
\author{Alberto Quaini\thanks{Corresponding author. I thank Anna Mikusheva for a fruitful exchange at SoFiE 2025 in Paris.}\\ Erasmus School of Economics, Erasmus University Rotterdam\\ P.O. Box 1738, 3000 DR Rotterdam, The Netherlands\\ \texttt{quaini@ese.eur.nl}}
\date{}
\maketitle
\begin{abstract}
Forecast comparisons inform investment and other economic decisions, but machine learning forecasts can be trained under objectives that differ from the relevant losses. Estimation error can then create spurious evidence of predictive superiority, even with out-of-sample evaluation. We establish feasible inference for differences in population evaluation risks in this setting. Building on locally robust estimation, we construct scores that preserve the risk target and remove first-order estimation sensitivity. Under geometric beta-mixing, moment conditions and simultaneous learning bounds, we establish Gaussian limits and consistent long-run covariance estimation for overlapping chronological training windows, allowing refitting at every evaluation date. Uniform extensions provide confidence coverage over classes of data-generating processes; Wald inference requires nondegenerate corrected-score covariance. For sparse linear forecasts and neural probability forecasts, we give explicit corrections and simultaneous learning bounds. Simulations with lasso and neural forecasts show substantial reductions in size distortion that bootstrapping uncorrected scores does not reproduce.
\end{abstract}

\section{Introduction}\label{sec:introduction}

Assessing whether a forecasting rule improves economic decisions requires
inference under the loss relevant to those decisions. This loss can differ
from the objective used to estimate the rule. A conditional mean estimated
by squared loss may be assessed under asymmetric costs \citep{Gneiting2011}.
Return forecasts trained under robust Huber objectives can be evaluated
by squared prediction error \citep{GuKellyXiu2020}.
Least-squares return forecasts can instead be judged by investment utility
\citep{CampbellThompson2008,WelchGoyal2008,RapachStraussZhou2010}.
Covariance forecasts chosen to minimize statistical estimation error may
likewise be assessed by portfolio utility \citep{FlemingKirbyOstdiek2001}.
Even with squared loss, persistent (nonvanishing) regularization penalties
can separate training from unpenalized evaluation
\citep{KellyMalamudZhou2024,DobribanWager2018}.
Evaluation focused on particular economic states can also differ from
unweighted training \citep{OdendahlRossiSekhposyan2023}.
Such choices have practical consequences; for instance, conventional forecast-error measures and
profits can rank forecasts differently \citep{LeitchTanner1991}, and
asymmetric loss changes assessments of forecast rationality
\citep{ElliottKomunjerTimmermann2005,ElliottKomunjerTimmermann2008}.

When evaluation risk is sensitive to estimation at first order, learning
error can create spurious evidence of predictive superiority even in
out-of-sample comparisons. We establish feasible inference for differences
in population evaluation risks when training objectives need not minimize
evaluation risk, allowing forecasts and corrections to be learned
repeatedly from overlapping past windows.

Our target compares the evaluation risks of population training solutions
and includes instrumented comparisons across economic states. It builds
on \citet{West1996} and the loss contrast of \citet{DieboldMariano1995}. It differs from comparisons with a fixed
estimation window, where estimation uncertainty remains part of the target
\citep{GiacominiWhite2006}, and from risk conditional on a realized training
sample or averaged over training samples of a fixed size
\citep{BatesHastieTibshirani2024}. Stability-based cross-validation
inference offers a different route to prediction-error assessment
\citep{BayleEtAl2020,AusternZhou2025}.

The leading estimation effect at a population training target $\eta_0$ is
$DR(\eta_0)[\widehat\eta-\eta_0]$, where $R(\eta)=\E\ell_t(\eta)$
is the evaluation risk. Matching objectives makes this derivative zero
at an interior training optimum; different losses can also share an
optimizer \citep{Gneiting2011}. \citet{EscancianoParra2026} establish
predictive-risk inference for machine learning forecasts when the expected
evaluation derivative is zero and learning is sufficiently fast. Their
analysis allows different losses but does not treat a nonzero expected
evaluation derivative. In that case, regularization bias and limited
learning accuracy can invalidate uncorrected inference. Classical
adjustments account for estimation effects through an asymptotic
description of the fitted parameters
\citep{West1996,WestMcCracken1998,McCracken2000}.
\citet{DemetrescuHanckHoga2026} use maxima over short rolling evaluation
windows to make estimation effects negligible and detect transient
predictive advantages. \citet{WilliamsonEtAl2023} compare optimal
population predictiveness, where optimizing the evaluation criterion
can itself remove first-order sensitivity. Our training targets need
not optimize that criterion.

We apply established orthogonal and locally robust moment constructions
\citep{Neyman1959,ChernozhukovEtAl2018,ChernozhukovEtAl2022LR}.
A representer relates each rule's evaluation derivative to its training
Hessian. Subtracting the training score in that direction preserves the
risk target and removes first-order sensitivity of the expected score.
Representer estimation follows \citet{ChernozhukovNeweySingh2022}.
The rules may differ in training objective, prediction class and
information set. These standard population arguments yield a mean-error bound involving
squared forecast error and the product of forecast and representer errors.
Correction also changes sampling variation: even coincident population
forecasts can have a nondegenerate corrected contrast. Related locally
robust methods address predicted outcomes \citep{EscancianoPuertaCuartas2026}
and functionals of conditional distributions \citep{EscancianoGazzei2026}.

Our main contribution concerns inference with repeated chronological
estimation under temporal dependence. An evaluation observation can enter
a later training window, linking subsequent fits to earlier evaluation
outcomes. Orthogonality controls population mean error; a separate argument
is needed to show that the centered score error from estimation is
asymptotically negligible. Under geometric beta-mixing, moment conditions
and simultaneous learning bounds, we establish asymptotic linearity for
fits based only on past data, allowing refitting at every evaluation date.
The rate conditions separately control mean error and centered score
error, with temporal separation entering the latter. A Bartlett covariance bound gives feasible Gaussian and Wald
inference under nondegenerate corrected-score long-run covariance.
Under uniform versions of these conditions, we obtain Gaussian
approximation and confidence coverage over classes of data-generating
processes without requiring covariance convergence. The number of refits enters through simultaneous learning
accuracy, with no additional algebraic factor in these rate conditions. No limiting distribution for the learners is required.

This analysis builds on orthogonal inference for dependent data.
\citet{SemenovaEtAl2023} leave out neighboring blocks when fitting
nuisance functions; their training sets can include earlier and later
blocks. \citet{CiganovicEtAl2026} use time-reversed auxiliary blocks
under time reversibility and conditional stability restrictions, with
a fixed number of folds. Our chronological argument permits overlapping
training windows and a growing number of refits, controlling
training--evaluation dependence through temporal gaps and beta-mixing.
Sequential influence-function methods provide a related precedent
\citep{LuedtkeVanDerLaan2016}.

For lasso, we adapt time-series estimation arguments \citep{WongLiTewari2020}
to obtain simultaneous forecast and correction rates over overlapping
windows, including under approximate sparsity. We control generated-response
error on the forecast training sample without the additional cross-fitting
used for nonlinear functionals by \citet{ChernozhukovNeweySingh2022}.
Neural event probabilities under cost-weighted evaluation have an analytic
correction. Their simultaneous learning bound separates approximation,
class complexity and an assumed empirical optimization gap.

Simulations compare raw and orthogonal tests and bootstrap versions of
each. Correction substantially reduces size distortion for cross-validated
lasso across dictionary sizes, training histories and refitting frequencies,
and for probability forecasts from rectified linear unit (ReLU) networks.
Bootstrapping raw evaluation scores does not reproduce these gains.
Learning diagnostics clarify the remaining distortion. Relative to the
raw test, size-adjusted power improves against some lasso alternatives but
falls at stronger negative alternatives in the neural design.

\subsection{Notation}

Write $v^{\T}$ for transpose. Unless specified otherwise,
$\norm{\cdot}$ is the Euclidean norm of vectors or Frobenius norm of
matrices; $\norm{\cdot}_{\rm op}$ denotes the operator norm.
For a random vector $V$,
$\norm V_p:=\{\E(\norm V^p)\}^{1/p}$.
For a probability law $P$, $L^2(P)$ is the Hilbert space of measurable
real functions with finite squared integral, identified when equal
$P$ almost surely. Its norm is
$\norm f_{L^2(P)}:=\{\int f^2\,dP\}^{1/2}$.
For functions of $X$, write $\norm f_2:=\norm f_{L^2(P_X)}$.
$D$ denotes a Fr\'echet derivative, with the direction displayed in
brackets when the derivative is applied. Constants $C,c>0$ may change between
displays and are uniform in the sample index. The notation
$a_n\lesssim b_n$ means $a_n\le Cb_n$, and $a_n\asymp b_n$
means both inequalities. We use $\op(1)$ and $\Op(1)$ for convergence
to zero and boundedness in probability.

Section~\ref{sec:population} defines the target, correction and local
error bounds. Section~\ref{sec:inference} establishes chronological
inference, and Section~\ref{sec:learner-summary} gives sufficient learning
conditions for two forecasting procedures. Technical assumptions are collected in
Appendix~\ref{sec:technical-conditions}. The supplement contains all
proofs, supporting results and full simulation designs.

\section{Population theory}\label{sec:population}

\subsection{Forecasting rules and risk targets}\label{sec:framework}

Let $Z_t=(Y_t,X_{1t},X_{2t},q_t)$, where $Y_t$ is the outcome at
date $t$, and the predictors $X_{1t},X_{2t}$ and instrument
$q_t\in\R^d$ are available at the forecast origin. The horizon and
$d$ are fixed. The instrument is prespecified, may use additional
information at the origin, and does not depend on the forecasting
parameters. Expectations below use the marginal law $P$.
For $j\in\{1,2\}$, write $\Ii_j:=\sigma(X_j)$.

Rule $j$ has parameters $\eta_j\in\mathcal D_j\subseteq\Hh_j$,
where $\Hh_j$ is a real separable Hilbert space, and an
$\Ii_j$-measurable prediction $F_j(\eta_j)$, evaluated at $X_{jt}$.
Parameterizations need not be injective or parameter spaces nested.
The training criterion $\rho_{j,t}$ defines
\begin{equation}\label{eq:training}
 Q_j(\eta_j):=\E\{\rho_{j,t}(\eta_j)\},\qquad
 \eta_{j,0}\in\arg\min_{\eta_j\in\mathcal D_j}Q_j(\eta_j).
\end{equation}
The score construction requires only $DQ_j(\eta_{j,0})=0$, as at an
interior local optimum of a differentiable $Q_j$.

Training need not minimize evaluation risk. For an evaluation loss
$\ell$, define, for $r\le d$ and $\eta:=(\eta_1,\eta_2)$,
\begin{equation}\label{eq:risks}
 R_{j,r}(\eta_j):=\E[q_{t,r}\ell\{Y_t,F_j(\eta_j)\}],\qquad
 M_r(\eta):=R_{1,r}(\eta_1)-R_{2,r}(\eta_2).
\end{equation}
Writing $M:=(M_r:r\le d)$ and
$\eta_0:=(\eta_{1,0},\eta_{2,0})$, we use the loss contrast of
\citet{DieboldMariano1995} and the population risk perspective of
\citet{West1996}:
\begin{equation}\label{eq:target}
 g_t(\eta):=q_t[\ell\{Y_t,F_1(\eta_1)\}
                         -\ell\{Y_t,F_2(\eta_2)\}],\qquad
 \theta_0:=M(\eta_0)=\E\{g_t(\eta_0)\}.
\end{equation}
The null $\theta_0=0$ states equality of the instrumented population
risks. For $d=1$ and $q_t=1$, these are ordinary expected losses.
Without an established risk ordering, the scalar alternative is two-sided.
A fixed benchmark requires no orthogonal correction of the type
constructed below; for a single rule's risk, omit the second loss
and correction.

The prediction classes
$\Fc_j:=\{F_j(\eta_j):\eta_j\in\mathcal D_j\}$ are identified
$P$ almost surely. Neither the information sets nor the prediction
classes need be nested; nesting of the former does not imply nesting
of the latter.

\subsection{Orthogonal scores and local error bounds}\label{sec:orthogonal}

Theorem~\ref{thm:population} and Proposition~\ref{prop:bounds} adapt
standard locally robust and Riesz correction arguments of
\citet{ChernozhukovEtAl2022LR} and
\citet{ChernozhukovNeweySingh2022} to provide the
score and error bounds for chronological inference. Population
expectations hold nuisance arguments fixed; parameter and representer
norms are the norms of $\Hh_j$.

Let $s_{j,t}(\eta_j)[v_j]$ be an integrable training score, linear
in $v_j\in\Hh_j$, satisfying
\begin{equation}\label{eq:scoreidentity}
 \E\{s_{j,t}(\eta_j)[v_j]\}=DQ_j(\eta_j)[v_j].
\end{equation}
Appendix Assumption~\ref{ass:population} specifies differentiability,
measurability and the first-order condition $DQ_j(\eta_{j,0})=0$.
With $\epsilon_1:=1$ and $\epsilon_2:=-1$, define
\[
 A_{j,0}(u_j,v_j):=D^2Q_j(\eta_{j,0})[u_j,v_j],
 \qquad G_{j,r}[u_j]:=\epsilon_jDR_{j,r}(\eta_{j,0})[u_j].
\]
Since $DM_r(\eta_0)[u]=\sum_jG_{j,r}[u_j]$, the raw moment is
orthogonal on the product parameter space exactly when every
$G_{j,r}=0$. A nonzero $G_{j,r}$ identifies the first-order
estimation effect to be removed.

\begin{assumption}[Existence of representers]\label{ass:representers}
For each $j,r$, there is $\alpha_{j,r,0}\in\Hh_j$ such that
\begin{equation}\label{eq:riesz}
 A_{j,0}(u_j,\alpha_{j,r,0})=G_{j,r}[u_j],
 \qquad u_j\in\Hh_j.
\end{equation}
\end{assumption}

Under Appendix Assumption~\ref{ass:population}, coercivity,
$A_{j,0}(v,v)\ge\kappa_j\norm v^2$ with
$\kappa_j>0$, suffices: the Riesz representation theorem gives a unique
solution with $\norm{\alpha_{j,r,0}}\le\kappa_j^{-1}\norm{G_{j,r}}$,
where the functional norm is the operator norm. Coercivity is not
required. If solutions are nonunique, fix one to which the estimator
converges; the corrected score and its covariance can depend on this
choice. When $G_{j,r}=0$, training and evaluation are locally aligned
for that component, and zero is an admissible representer.

Set $\alpha_j:=(\alpha_{j,r}:r\le d)$ and
$\nu:=(\eta_1,\eta_2,\alpha_1,\alpha_2)$, with population value
$\nu_0$. The corrected score and moment are
\begin{equation}\label{eq:corrected}
 \phi_{t,r}(\nu):=g_{t,r}(\eta)
                    -\sum_{j=1}^2s_{j,t}(\eta_j)[\alpha_{j,r}],
 \qquad
 \Psi(\theta,\nu):=\E\{\phi_t(\nu)\}-\theta.
\end{equation}

\begin{theorem}[Representers and orthogonality]\label{thm:population}
Under Assumption~\ref{ass:representers} and Appendix
Assumption~\ref{ass:population},
\begin{equation}\label{eq:orthogonality}
 \Psi(\theta_0,\nu_0)=0,\qquad
 D_\nu\Psi(\theta_0,\nu_0)[u,b]=0
\end{equation}
for all directions $u=(u_1,u_2)$ and $b=(b_{j,r}:j\le2,r\le d)$
with $u_j,b_{j,r}\in\Hh_j$.
\end{theorem}

The training first-order conditions preserve the target and eliminate
sensitivity to representer estimation; \eqref{eq:riesz} cancels
first-order sensitivity to forecast estimation. The evaluation risk
itself may still have a nonzero derivative. The next result quantifies
the remaining mean error and the $L^2$ score perturbation.

\begin{proposition}[Local mean and score bounds]\label{prop:bounds}
Under Assumption~\ref{ass:representers} and Appendix
Assumptions~\ref{ass:population}--\ref{ass:local-technical},
consider a neighborhood within the specified local domains on which
$\norm{\eta_j-\eta_{j,0}}\le a_j$ and
$\{\sum_{r=1}^d\norm{\alpha_{j,r}-\alpha_{j,r,0}}^2\}^{1/2}\le c_j$
for $j\in\{1,2\}$. Uniformly over this neighborhood,
\begin{equation}\label{eq:biasbound}
 \norm{\E\{\phi_t(\nu)\}-\theta_0}
 \lesssim\sum_j(a_j^2+a_jc_j),\qquad
 \norm{\phi_t(\nu)-\phi_t(\nu_0)}_2
 \lesssim\sum_j(a_j+c_j).
\end{equation}
The constants do not depend on the neighborhood radii.
\end{proposition}

With regularity constants uniform in $n$, learning bounds
$a_{j,n}$ and $c_{j,n}$ yield the mean and $L^2$ error scales
\begin{equation}\label{eq:rates}
 B_n:=\sum_j(a_{j,n}^2+a_{j,n}c_{j,n}),\qquad
 \rho_n:=\sum_j(a_{j,n}+c_{j,n}).
\end{equation}
The mean bound uses population smoothness; the $L^2$ bound requires
additional score moments, or may be verified directly.
Section~\ref{sec:inference} establishes how these
bounds support Gaussian inference and long-run covariance estimation
with overlapping training windows and refitting at every evaluation date.

\subsection{What the correction changes}
\label{sec:consequences}

The correction removes the linear effect of forecast estimation on the
mean score. To make the comparison explicit, write
$u_j:=\eta_j-\eta_{j,0}$ and
$G_j[u_j]:=(G_{j,r}[u_j]:r\le d)$. On local neighborhoods where
each $DR_{j,r}$ is Lipschitz, Taylor's formula gives
\begin{equation}\label{eq:raw-local-mean}
 \E\{g_t(\eta)\}-\theta_0
   =\sum_jG_j[u_j]+r_g(\eta),\qquad
 \norm{r_g(\eta)}\lesssim\sum_j\norm{u_j}^2.
\end{equation}
Under Appendix Assumption~\ref{ass:population} alone, the remainder
is $o(\sum_j\norm{u_j})$; the quadratic bound requires the stated
additional smoothness. Together with \eqref{eq:scorelocal1},
\eqref{eq:raw-local-mean} gives, uniformly for $\norm{u_j}\le a_j$,
\begin{equation}\label{eq:raw-local-bounds}
 \norm{\E\{g_t(\eta)\}-\theta_0}
   \lesssim\sum_j(\norm{G_j}a_j+a_j^2),\qquad
 \norm{g_t(\eta)-g_t(\eta_0)}_2\lesssim\sum_ja_j.
\end{equation}
These are standard expansion bounds, consistent with the estimation
effects studied by \citet{West1996} and the zero-derivative analysis
of \citet{EscancianoParra2026}.

With uniform local constants and bounded evaluation derivatives,
$\sqrt n\sum_ja_{j,n}\to0$ suffices to make the raw mean error negligible.
For the corrected mean error, it suffices that
$\sqrt n\sum_j(a_{j,n}^2+a_{j,n}c_{j,n})\to0$, which holds when both
errors are $o(n^{-1/4})$. Vanishing derivatives or cancellation can
weaken the raw requirement. Correction also adds representer error to
$\rho_n$; Section~\ref{sec:inference} controls its contribution to the
centered score remainder and covariance estimation.

The correction also changes sampling variation. Remove the contrast
signs from the selected representers by setting
$a_{j,r}:=\epsilon_j\alpha_{j,r,0}$. Then
$A_{j,0}(u,a_{j,r})=DR_{j,r}(\eta_{j,0})[u]$, and the population score is
\begin{equation}\label{eq:correction-variance}
 \phi_{t,r}(\nu_0)
 =g_{t,r}(\eta_0)-s_{1,t}(\eta_{1,0})[a_{1,r}]
                         +s_{2,t}(\eta_{2,0})[a_{2,r}].
\end{equation}
Its long-run covariance includes both corrections and their
cross-covariances with the raw loss contrast, and need not be larger
or smaller than the raw covariance. If
$F_1(\eta_{1,0})=F_2(\eta_{2,0})$ almost surely, the raw contrast
vanishes, and the corrected component vanishes almost surely exactly when
\[
 s_{1,t}(\eta_{1,0})[a_{1,r}]
   =s_{2,t}(\eta_{2,0})[a_{2,r}]\quad P\text{ almost surely}.
\]
Example~\ref{ex:asymmetric-evaluation} gives a failure of this
cancellation with positive corrected variance. Conversely, a nonzero
evaluation derivative does not ensure nondegenerate corrected-score
covariance. Ordinary Wald inference requires the latter condition
separately. Supplementary Section~\external{sec:degeneracy} records
the covariance identity and a failure of ordinary studentization at a
degenerate score.

\subsection{Aligned training and evaluation}
\label{sec:alignment}

No correction is needed for local orthogonality precisely when the
raw moment has zero derivative on the product parameter space:
\begin{equation}\label{eq:zero-correction}
 \begin{aligned}
 DM(\eta_0)=0
 &\quad\Longleftrightarrow\quad
 DR_{j,r}(\eta_{j,0})=0\ \text{for every }j,r\\
 &\quad\Longleftrightarrow\quad
 \alpha_{j,r,0}=0\ \text{solves \eqref{eq:riesz} for every }j,r.
 \end{aligned}
\end{equation}
A sufficient condition, applied separately to each rule and component,
is $R_{j,r}=c_{j,r}Q_j+C_{j,r}$ locally, where $c_{j,r}>0$ and
$C_{j,r}$ do not depend on $\eta_j$. The training first-order
condition then gives $DR_{j,r}(\eta_{j,0})=0$. Different losses
can also share a stationary target; equality of their formulas is
unnecessary. Conversely, a nonconstant instrument can prevent
alignment even under the same pointwise training and evaluation loss,
as Example~\ref{ex:state-evaluation} shows.

Setting the representers identically to zero under
\eqref{eq:zero-correction} gives $\phi_t=g_t$ and $c_{j,n}=0$.
Under the local conditions of
Proposition~\ref{prop:bounds}, the error scales reduce to
\[
 B_n=\sum_ja_{j,n}^2,\qquad \rho_n=\sum_ja_{j,n}.
\]
This is the zero-derivative, fast-learning mechanism studied by
\citet{EscancianoParra2026}. Their stochastic equicontinuity and
learning conditions control the remaining estimation error; here the
conditions in Section~\ref{sec:inference}
control it for chronological fits. Alignment alone does not ensure
valid Wald inference: if the
population forecasts also coincide, then $\phi_t(\nu_0)=0$ almost
surely and the corrected-score covariance is degenerate.

\section{Estimation and inference}\label{sec:inference}

\subsection{Chronological fits and estimated scores}\label{sec:dependence}

To study the equality null and shrinking local alternatives within one
asymptotic framework, the remaining results allow distributions, parameter
spaces, criteria, prediction classes and targets to vary with $n$,
while the forecast horizon and $d$ remain fixed. Write $\phi_{t,n}(\nu)=\phi_n(Z_{t,n};\nu)$ for the score,
$P_n$ for the stationary marginal law and
\[
 \theta_n:=\E\{\phi_{t,n}(\nu_{0n})\}=M_n(\eta_{0n}),\qquad
 \xi_{t,n}:=\phi_{t,n}(\nu_{0n})-\theta_n.
\]
Population identification is required at every $n$. The constructions
above provide it. Subscripts $n$ on individual observations and
score maps are suppressed where unambiguous.

We use a strictly stationary observation array with geometric absolute
regularity: $\beta_n(h)\le C_\beta e^{-c_\beta h}$, uniformly in $n$.
The precise dependence definition, observation-space conditions and
oracle moment bound are in Appendix Assumption~\ref{ass:array}
and its preceding discussion.

Define
$\Omega_n:=\sum_{h\in\mathbb Z}\cov(\xi_{0,n},\xi_{h,n})$.
Absolute convergence of this series and positive semidefiniteness
follow from Lemma~\external{lem:moments} and the covariance-sum
identity in Supplementary Section~\external{sec:degeneracy}.
A limit as $n\to\infty$ and positive definiteness are not imposed here.

\begin{assumption}[Chronological fits]\label{ass:chronology}
The dates $1,\ldots,n$ are partitioned into deterministic consecutive
nonempty blocks $I_k$, with first dates $t_k$ and $K_n\le n$.
Let
$\Ff_{t,n}:=\sigma(Z_{s,n}:s\le t)\vee\mathcal A_n$,
where $\mathcal A_n$ is fitting randomness independent of the entire
observation array. For an integer $b_n\ge1$, each
fit $\widehat\nu_k$ is $\Ff_{t_k-b_n-1,n}$-measurable.
For a forecast horizon $h$, require also $b_n+1\ge h$, so fitting
uses observations available at the first forecast origin.
Training windows may overlap.
\end{assumption}

\begin{assumption}[Localization]\label{ass:learning}
There are deterministic
sets $\mathcal N_n$ containing $\nu_{0n}$, and deterministic
$B_n,\rho_n\to0$, such that
\begin{equation}\label{eq:localization}
 \pr\{\widehat\nu_k\in\mathcal N_n\text{ for every }k\}\to1.
\end{equation}
For $\Delta_{t,n}(\nu):=\phi_{t,n}(\nu)-\phi_{t,n}(\nu_{0n})$,
\begin{equation}\label{eq:meanlocal}
 \sup_{\nu\in\mathcal N_n}\norm{\E\Delta_{t,n}(\nu)}\le CB_n,
 \qquad
 \sup_{\nu\in\mathcal N_n}\norm{\Delta_{t,n}(\nu)}_2\le C\rho_n.
\end{equation}
Expectations in \eqref{eq:meanlocal} hold their
argument fixed.
\end{assumption}

The auxiliary randomness $\mathcal A_n$ covers random seeds for initialization, minibatch
ordering and validation splits; it is trivial for deterministic fits.
The buffer $b_n$ separates each fit from its evaluation block, controlling
residual training--evaluation dependence while respecting forecast
availability. Localization is standard in double machine learning
\citep{ChernozhukovEtAl2018}; here it is required simultaneously over
chronological fits. Uniform versions of Proposition~\ref{prop:bounds}, together with
simultaneous learning bounds, supply \eqref{eq:meanlocal}.
The oracle moment condition in Appendix Assumption~\ref{ass:array} is additional; it does not follow from
an $L^2$ learning rate.

For a vector sequence $x_1,\ldots,x_n$, let
$\bar x:=n^{-1}\sum_tx_t$, $y_t:=x_t-\bar x$, and define
\begin{align}
 \widehat\Gamma_x(h)&:=n^{-1}\sum_{t=h+1}^n y_ty_{t-h}^{\T},\qquad
                      0\le h\le L_n,\nonumber\\
 \mathcal B_n(x)&:=\widehat\Gamma_x(0)+
       \sum_{h=1}^{L_n}\left(1-\frac h{L_n+1}\right)
        \{\widehat\Gamma_x(h)+\widehat\Gamma_x(h)^{\T}\}.
        \label{eq:bartlett}
\end{align}
Here $1\le L_n<n$ is deterministic. This estimator is positive
semidefinite; its block-sum representation and covariance bounds are
proved in Supplementary Lemma~\external{lem:haccompare}.

Use the appropriate population score from Section~\ref{sec:population}
and set
\[
 \widehat\phi_t:=\phi_t(\widehat\nu_{k(t)}),\qquad
 \widehat\theta_n:=n^{-1}\sum_t\widehat\phi_t,\qquad
 \widehat\Omega_n:=\mathcal B_n(\widehat\phi).
\]
The score includes both estimation corrections. Its long-run covariance
therefore includes their variances and their covariances with the raw
loss difference. The same construction covers common and different
training objectives.

\subsection{Gaussian limits and feasible inference}

When the corrected score has a positive definite limiting long-run
covariance, the estimated risk contrast is asymptotically Gaussian and
its estimated covariance delivers feasible Wald inference.

\begin{theorem}[Gaussian inference]\label{thm:ordinary}
Suppose Assumptions~\ref{ass:chronology}--\ref{ass:learning} and Appendix Assumption~\ref{ass:array} hold,
$\Omega_n\to\Omega$ with $\Omega$ positive definite, and
\begin{equation}\label{eq:ordinaryrates}
 n\beta_n(b_n)\to0,\qquad
 \sqrt{b_n+1}\rho_n\to0,\qquad \sqrt n B_n\to0.
\end{equation}
Then
\begin{equation}\label{eq:linear}
 \sqrt n(\widehat\theta_n-\theta_n)
      =n^{-1/2}\sum_t\xi_{t,n}+\op(1)
                   \Longrightarrow N_d(0,\Omega).
\end{equation}
\end{theorem}

\begin{corollary}[Covariance estimation and Wald inference]\label{cor:wald}
Under the assumptions of Theorem~\ref{thm:ordinary}, suppose also
$L_n\to\infty$, $L_n/n\to0$ and $\sqrt{L_n}\rho_n\to0$. Then $\widehat\Omega_n\to\Omega$ in probability. Define
$W_n:=n\widehat\theta_n^{\T}\widehat\Omega_n^{-1}\widehat\theta_n$,
with value zero on the event that the inverse does not exist.
For $\sqrt n\theta_n\to h$,
\[
 W_n\Longrightarrow\chi_d^2(h^{\T}\Omega^{-1}h).
\]
In particular, the test using the $1-\alpha$ quantile of
$\chi_d^2$ has asymptotic size $\alpha$ under $\theta_n=0$.
If $\norm{\sqrt n\theta_n}\to\infty$, its power tends to one.
\end{corollary}

The noncentrality parameter $h^{\T}\Omega^{-1}h$ determines local power:
it measures the local risk difference relative to corrected-score
variability. Under the maintained conditions, the test is consistent
when the distance from the null decreases more slowly than $n^{-1/2}$.

\subsection{Uniform Gaussian approximation and coverage}

\begin{proposition}[Uniform Gaussian approximation and coverage]\label{prop:uniform}
Fix $d$, and let $\mathfrak P_n$ be a nonempty class of laws and fitted
procedures. Suppose the assumptions and rate conditions of
Theorem~\ref{thm:ordinary} and Corollary~\ref{cor:wald} hold uniformly
over $P\in\mathfrak P_n$, with common moment and mixing constants,
uniform local mean and $L^2$ bounds, and
\begin{equation}\label{eq:uniform-localization}
 \sup_{P\in\mathfrak P_n}
 \pr_P\{\widehat\nu_k\notin\mathcal N_{n,P}
                       \text{ for some }k\}\longrightarrow0.
\end{equation}
Replace covariance convergence by
$\inf_{P\in\mathfrak P_n}\lambda_{\min}(\Omega_{n,P})\ge c>0$
eventually. Let $\widehat L_n$ be the lower triangular Cholesky
factor of $\widehat\Omega_n$, with positive diagonal, and put
$T_{n,P}=\widehat L_n^{-1}\sqrt n(\widehat\theta_n-\theta_{n,P})$
and $Q_{n,P}=\norm{T_{n,P}}^2$. Define both as zero when
$\widehat\Omega_n$ is singular. Then
\begin{equation}\label{eq:uniform-gaussian}
 \sup_{\substack{P\in\mathfrak P_n\\x\in\R^d}}
 \left|\pr_P(T_{n,P}\le x)-\Phi_d(x)\right|\to0,
 \qquad
 \sup_{\substack{P\in\mathfrak P_n\\u\in\R}}
 \left|\pr_P(Q_{n,P}\le u)-F_{\chi_d^2}(u)\right|\to0,
\end{equation}
where inequalities are coordinatewise and $\Phi_d$ is the distribution
function of $N_d(0,I_d)$. For every fixed $\alpha\in(0,1)$, the region
\[
 \mathcal C_{n,1-\alpha}:=
 \left\{\vartheta\in\mathbb R^d:
 n(\widehat\theta_n-\vartheta)^{\T}\widehat\Omega_n^{-1}
       (\widehat\theta_n-\vartheta)
       \le\chi_{d,1-\alpha}^2\right\},
\]
defined as $\mathbb R^d$ on the singular event, satisfies
\begin{equation}\label{eq:uniform-coverage}
 \sup_{P\in\mathfrak P_n}
 \left|\pr_P\{\theta_{n,P}\in\mathcal C_{n,1-\alpha}\}
                       -(1-\alpha)\right|\longrightarrow0.
\end{equation}
\end{proposition}

These conclusions require uniform learning bounds over the stated law
class; pointwise learner consistency is insufficient. The dimension
remains fixed, and the covariance lower bound excludes laws approaching
corrected-score degeneracy.

\subsection{Comparison with the uncorrected statistic}
\label{sec:raw-comparison}

The same chronological argument identifies what orthogonalization removes.
Let $\widehat g_t:=g_t(\widehat\eta_{k(t)})$,
$\widehat\theta_n^g:=n^{-1}\sum_t\widehat g_t$, and define
\[
 \xi^g_{t,n}:=g_t(\eta_{0n})-\theta_n,\qquad
 \Delta_n:=\frac1{\sqrt n}\sum_k|I_k|
              \{M_n(\widehat\eta_k)-\theta_n\}.
\]
Here $\Delta_n$ accumulates the population mean errors of the fitted
forecasts, with each fit held fixed inside $M_n$.

\begin{proposition}[Uncorrected inference and estimation drift]\label{prop:raw-comparison}
Suppose Assumption~\ref{ass:chronology} holds for $\widehat\eta_k$
and Appendix Assumption~\ref{ass:array} holds with $\xi^g$ in place
of $\xi$. Let
$\Omega_n^g:=\sum_{h\in\mathbb Z}\cov(\xi^g_{0,n},\xi^g_{h,n})
\to\Omega^g$, with $\Omega^g$ positive definite.
Suppose there are deterministic sets $\mathcal E_n$ containing
$\eta_{0n}$ and $r_n\to0$ such that
\[
 \pr\{\widehat\eta_k\in\mathcal E_n\text{ for every }k\}\to1,
 \qquad
 \sup_{\eta\in\mathcal E_n}
   \norm{g_t(\eta)-g_t(\eta_{0n})}_2\le Cr_n.
\]
If $n\beta_n(b_n)\to0$ and $\sqrt{b_n+1}\,r_n\to0$, then
\begin{equation}\label{eq:raw-expansion}
 \sqrt n(\widehat\theta_n^g-\theta_n)
       =n^{-1/2}\sum_t\xi^g_{t,n}+\Delta_n+\op(1).
\end{equation}
If also $L_n\to\infty$, $L_n/n\to0$ and
$\sqrt{L_n}r_n\to0$, then
$\widehat\Omega_n^g:=\mathcal B_n(\widehat g)\to\Omega^g$
in probability. Define
$W_n^g:=n(\widehat\theta_n^g)^{\T}(\widehat\Omega_n^g)^{-1}
\widehat\theta_n^g$, with value zero when the inverse does not exist.
If $\sqrt n\theta_n\to h$ and $\Delta_n\to\delta$ in probability
for a deterministic $\delta\in\mathbb R^d$, then
\begin{equation}\label{eq:raw-wald}
 W_n^g\Longrightarrow
 \chi_d^2\{(h+\delta)^{\T}(\Omega^g)^{-1}(h+\delta)\}.
\end{equation}
In particular, under the null $\theta_n=0$, put
$\lambda_\delta:=\delta^{\T}(\Omega^g)^{-1}\delta$.
For every $\alpha\in(0,1)$,
\begin{equation}\label{eq:raw-size}
 \pr\{W_n^g>\chi_{d,1-\alpha}^2\}
 \longrightarrow 1-F_{d,\lambda_\delta}(\chi_{d,1-\alpha}^2)
 \begin{cases}
  =\alpha,&\delta=0,\\
  >\alpha,&\delta\ne0,
 \end{cases}
\end{equation}
where $F_{d,\lambda}$ is the distribution function of $\chi_d^2(\lambda)$.
\end{proposition}

The raw mean error is generally first order in forecast error, whereas
Proposition~\ref{prop:bounds} gives a squared-error and product-error bound
after correction. A first-order rate bound can nevertheless be conservative
when evaluation derivatives vanish or estimation errors cancel.
A random limiting drift requires its joint limit with the oracle sum;
the Wald limit need not be noncentral chi-squared.
Estimation-aware inference can also retain the raw estimate when an
influence expansion for the fitted rules is available
\citep{West1996,WestMcCracken1998}. Coincident population forecasts give
zero raw oracle covariance, a case outside this positive-covariance comparison.

\subsection{Accuracy, covariance and implementation}

The learning restrictions in Theorem~\ref{thm:ordinary} control two
errors: $\sqrt nB_n\to0$ makes the mean score error negligible, while
$\sqrt{b_n+1}\rho_n\to0$ controls the centered score-replacement error.
Repeated fitting makes this step different from independent sample
splitting: an evaluation observation may enter later training windows,
so conditioning on all fitted rules does not separate estimation from
evaluation. Lemma~\external{lem:replacement} instead centers each
truncated score error given the information available $b_n+1$ dates
earlier. Within each residue class modulo $b_n+1$, these centered
terms are pairwise uncorrelated. Cauchy--Schwarz across the classes
gives a leading variance bound of order $(b_n+1)\rho_n^2$ for their
normalized sum. Truncation and absolute regularity control the remaining
terms. This explains the factor $\sqrt{b_n+1}$: the argument groups
evaluation dates by temporal separation, not by fitted rule. It permits
$K_n=n$; the number of fits enters through simultaneous localization,
with no additional algebraic multiplier in \eqref{eq:ordinaryrates}.

The gap makes observations used for evaluation sufficiently separated
from fitting: under Appendix Assumption~\ref{ass:array}, choosing
$b_n=h+\lceil C_b\log(en)\rceil$ with $c_\beta C_b>1$ ensures
$n\beta_n(b_n)\to0$. This is a sufficient scheme with an explicit
loss of recent training observations.

The Bartlett block-sum representation bounds covariance replacement by
$\Op(\sqrt{L_n}\rho_n+L_n\rho_n^2)$, so long-run covariance estimation
adds $\sqrt{L_n}\rho_n\to0$. For example,
$L_n=\lceil n^{1/4}\rceil$ is admissible when $n^{1/8}\rho_n\to0$.
The constants governing dependence and score moments must remain
uniform, including when the predictor dictionary changes with $n$.
The mixing bound concerns the complete observation vector; increasing
the number of lags need not preserve it. Orthogonality alone supplies
none of these probability bounds.

Implementation follows the same sequence for the two procedures in
Section~\ref{sec:learner-summary}. Fix the evaluation loss, instruments,
forecasting procedures, estimation windows and gap before examining
evaluation outcomes. On each training window, estimate the forecasting
rules and their representers. Form the entire corrected score on the
subsequent evaluation block and average those scores. Apply the Bartlett
covariance estimator to that same sequence, retaining covariance between
the loss difference and both corrections. For a scalar risk contrast,
an asymptotic $1-\alpha$ confidence interval is
\[
 \widehat\theta_n\ \pm\ z_{1-\alpha/2}
                         \sqrt{\widehat\Omega_n/n}.
\]
The corresponding test of a prespecified value $\theta^\star$ uses
$\sqrt n(\widehat\theta_n-\theta^\star)/\sqrt{\widehat\Omega_n}$.
These conclusions require the positive definite long-run covariance
in Theorem~\ref{thm:ordinary}; a nonzero evaluation derivative does not
ensure this condition. Supplementary Example~\external{ex:failure} shows 
that ordinary studentization can fail at a degenerate covariance
even when the quadratic mean-error rate holds.

For a state indicator $q_s$ with $\E q_s>0$, a conditional risk
difference is $\theta_s/\E q_s$. Estimating the denominator changes
the influence score and its covariance; Supplementary
Section~\external{sec:states} gives the corresponding normalization.

\section{Forecasting implementations and feasible corrections}
\label{sec:learner-summary}\label{sec:score-examples}

We now construct the scores and verify simultaneous learning bounds for
two forecasting procedures: sparse linear forecasts with an estimated
correction, and neural probability forecasts with an analytic correction.
For each evaluation block $I_k$, use the $T_n$ observations in
$J_k:=\{t_k-b_n-T_n,\ldots,t_k-b_n-1\}$ and write
$\E_kv:=T_n^{-1}\sum_{t\in J_k}v_t$. The windows may overlap and
$K_n=n$ permits refitting at every evaluation date.

Both procedures target squared-loss minimizers. Using squared
loss to define the population training score, the representer and
corrected score for an unrestricted conditional mean
$m_j:=\E(Y\mid X_j)$ are
\begin{align}
 \alpha_{jr,0}(X_j)
 &=\epsilon_j\E\{q_r\dot\ell(Y,m_j)\mid X_j\},
 \qquad \dot\ell(y,z):=\partial_z\ell(y,z),\label{eq:mean-representer}\\
 \phi_r(f,\alpha)&:=q_r\{\ell(Y,f_1)-\ell(Y,f_2)\}
                +\sum_j\alpha_{jr}(Y-f_j).\label{eq:mean-score}
\end{align}
Here $f_j$ and $\alpha_{jr}$ are functions of $X_j$. For projections
on a dictionary, the representer instead satisfies
\eqref{eq:projection-representer} below. Supplementary
Proposition~\external{prop:mean-correction}
and Corollary~\external{cor:projection} give the mean-error bounds
under Appendix Assumption~\ref{ass:mean-technical}.

\subsection{Sparse linear forecasts and estimated corrections}
\label{ex:lasso-implementation}

\begin{example}[Mean forecasts under asymmetric evaluation]
\label{ex:asymmetric-evaluation}\label{ex:intro-variance}
For fixed $\kappa\ge0$, let
\begin{equation}\label{eq:asymmetric-loss}
 \ell_\kappa(y,z):=(y-z)^2+\kappa(z-y)_+^2,
 \qquad x_+:=\max(x,0).
\end{equation}
This asymmetric least-squares loss \citep{NeweyPowell1987} makes
overprediction $1+\kappa$ times as costly as equally large
underprediction; see also \citet{ElliottKomunjerTimmermann2008}.
Its derivative is $\dot\ell_\kappa(y,z)=2(z-y)+2\kappa(z-y)_+$.
For $q=1$, the mean representer is
$\alpha_{j,0}=2\epsilon_j\kappa\E\{(m_j-Y)_+\mid X_j\}$,
which vanishes at $\kappa=0$ and is nonzero for $\kappa>0$ when
the conditional outcome is nondegenerate.

Coincident training means can nevertheless have a nondegenerate
corrected contrast. Let $W$ be uniform on $\{0,1\}$, let $S$ be
an independent symmetric sign, and set $Y:=\sigma(W)S$, with
distinct positive values of $\sigma(0)$ and $\sigma(1)$. Rule 1 uses
no predictors and rule 2 uses $W$. Both have training mean zero, but
their corrected contrast is
\[
 \phi_t=\kappa Y_t\{\E\sigma(W)-\sigma(W_t)\},\qquad
 \E\phi_t=0,\qquad \var(\phi_t)>0\quad(\kappa>0).
\]
The raw contrast is identically zero. Independent observations therefore
give positive corrected long-run variance; Supplementary
Section~\external{sec:degeneracy} gives dependent versions.
\end{example}

\begin{example}[Linear projections evaluated in a selected state]
\label{ex:state-evaluation}
Let $f_{j,0}:=X_j^{\T}\beta_{j,0}$ be the least-squares projection
on a deterministic dictionary $X_j\in\R^{p_n}$, allowing
misspecification, and let $\Sigma_j:=\E(X_jX_j^{\T})$ be positive
definite. For bounded scalar $q$ and a differentiable evaluation loss,
the coefficient representer solves
\begin{equation}\label{eq:projection-representer}
 \Sigma_j\alpha_{j,0}
   =\epsilon_j\E\{qX_j\dot\ell(Y,f_{j,0})\}.
\end{equation}
Use $f_j=X_j^{\T}\beta_j$ and $X_j^{\T}\alpha_j$ in
\eqref{eq:mean-score}. For a prespecified state $A$ observable at the
forecast origin, $q:=\ind\{A\}$ with $\pr(A)>0$ and $\ell=\ell_0$
give risk $\pr(A)\E\{(Y-f_j)^2\mid A\}$. The training normal
equations need not eliminate the weighted moment
$2\epsilon_j\E\{qX_j(f_{j,0}-Y)\}$, as when the dictionary omits
state-dependent interactions. If $f_{j,0}=m_j$ and $q$ is measurable
in $X_j$, the correction is zero.

For this state-weighted squared-loss comparison, our technical contribution
is to obtain simultaneous rates for the projection and its estimated
correction across overlapping chronological training windows
(Corollary~\ref{cor:lasso}). Both fits use the same eligible past sample.
These joint bounds verify the learning requirements for feasible inference
on the population state-weighted risk difference.
\end{example}

For evaluation loss $\ell_\kappa$ and a scalar instrument $q$, estimate
the projection and its representer on the same window by
\begin{equation}\label{eq:lasso-fit}
 \widehat\beta_{jk}\in\arg\min_\beta
 \{\tfrac12\E_k(Y-X_j^{\T}\beta)^2+\lambda_n\norm{\beta}_1\},
\end{equation}
\begin{equation}\label{eq:lasso-representer}
 \widehat H_{jt,k}:=\epsilon_jq_t\dot\ell_\kappa
          (Y_t,X_{jt}^{\T}\widehat\beta_{jk}),
 \qquad
 \widehat\alpha_{jk}\in\arg\min_\alpha
 \{\tfrac12\E_k(\widehat H_{j,k}-X_j^{\T}\alpha)^2
                         +\lambda_n\norm{\alpha}_1\}.
\end{equation}
Every coefficient, including an intercept if present, is penalized.
Let $s_{f,n},s_{a,n}$ bound the two population sparsities and set
\begin{equation}\label{eq:lasso-rate}
 r_n:=\left\{\frac{\log(enK_nT_n)\log(ep_nK_nn)}{T_n}\right\}^{1/2},
 \qquad \lambda_n:=C_\lambda r_n.
\end{equation}

\begin{corollary}[Forecast and correction estimation on the same window]
\label{cor:lasso}
Under Appendix Assumption~\ref{ass:lasso-technical}, define
\[
 a_n:=\sqrt{s_{f,n}}r_n,\qquad
 c_n:=(\sqrt{s_{f,n}}+\sqrt{s_{a,n}})r_n.
\]
With probability tending to one, simultaneously over rules and windows,
\[
 \norm{\widehat\beta_{jk}-\beta_{j,0}}\le Ca_n,\qquad
 \norm{\widehat\alpha_{jk}-\alpha_{j,0}}\le Cc_n.
\]
The localization bounds hold with $B_n=C(a_n^2+a_nc_n)$ and
$\rho_n=C(a_n+c_n)$. Theorem~\ref{thm:ordinary} and
Corollary~\ref{cor:wald} give feasible inference under their remaining
covariance, chronology, bandwidth and rate conditions.
\end{corollary}

Corollary~\ref{cor:lasso} jointly controls forecast and correction
estimation across overlapping chronological training windows.
\citet{WongLiTewari2020} establish time-series lasso bounds, while
\citet{ChernozhukovNeweySingh2022} use additional cross-fitting
to estimate derivatives of nonlinear regression functionals. We control
the generated-response error on the forecast training sample, without
additional splitting. Combined with uniform concentration over overlapping
windows, this yields simultaneous rates sufficient for chronological
Gaussian and Wald inference. The correction rate $c_n$ includes forecast
error because its response is estimated; forecast sparsity does not imply
representer sparsity. The mean-error condition is
$\sqrt n(s_{f,n}+\sqrt{s_{f,n}s_{a,n}})r_n^2\to0$.
Supplementary Proposition~\external{prop:lasso-approximate} extends
these bounds to approximately sparse forecasts and corrections in
$\ell_q$ balls with exponents in $(0,1)$, allowing dense coefficients without changing the targets.
The remaining inference conditions still apply; the deterministic
calibration in \eqref{eq:lasso-rate} does not justify cross-validated penalties.

\subsection{Neural probability forecasts and analytic corrections}
\label{ex:neural-implementation}

\begin{example}[Event probabilities under cost-weighted evaluation]
\label{ex:probability-evaluation}
For binary $Y$, squared-loss training targets
$p_{j,0}(X_j):=\pr(Y=1\mid X_j)$. With $q=1$ and a fixed cost-weight
parameter $\tau\ge0$, evaluate this probability under
\begin{equation}\label{eq:probability-evaluation}
 \ell_\tau(y,p):=(1+\tau y)(y-p)^2,\qquad \tau\ge0.
\end{equation}
This is an outcome-weighted squared loss as in
\citet{Gneiting2011}, with event weight
$1+\tau$. The conditional minimizer is
$(1+\tau)p_{j,0}/(1+\tau p_{j,0})$,
which differs from the training probability when $\tau>0$ and
$0<p_{j,0}<1$. The target remains the evaluation risk of $p_{j,0}$.
Equation~\eqref{eq:mean-representer} yields
\begin{equation}\label{eq:nn-representer}
 A_\tau(p):=-2\tau p(1-p),\qquad
 \alpha_{j,0}=\epsilon_jA_\tau(p_{j,0}).
\end{equation}
Writing $p_{jt}:=p_j(X_{jt})$, the corrected score is
\begin{equation}\label{eq:nn-score}
 H_\tau(y,p):=\ell_\tau(y,p)+A_\tau(p)(y-p),
 \qquad
 \phi_t(p):=H_\tau(Y_t,p_{1t})-H_\tau(Y_t,p_{2t}).
\end{equation}
At $\tau=0$, the correction is zero and this is the Brier-loss contrast.
\end{example}

Fit $\widehat p_{jk}$ by squared loss over a deterministic neural class
$\mathcal P_{j,n}$ with outputs in $[0,1]$, and insert it in both terms
of $H_\tau$. The class approximates the conditional probability;
it does not define a fixed misspecified target. The correction needs
neither an auxiliary regression nor a Hessian with respect to network weights.
Supplementary Proposition~\external{prop:nn-correction} gives
$B_n=Ca_n^2$ and $\rho_n=Ca_n$ for simultaneous $L^2$ prediction
error $a_n$.

Appendix Assumption~\ref{ass:nn-technical} defines the approximation
error $d_n$, log covering number
$H_n:=\max_j\log N(T_n^{-2},\mathcal P_{j,n},\norm{\cdot}_\infty)$,
and simultaneous empirical optimization gap $\varepsilon_n$.
Its block length $v_n:=\lceil C_v\log(enK_nT_n)\rceil$ accounts
for geometric mixing. Neural learning and inference are developed
under independent sampling by \citet{FarrellLiangMisra2021};
\citet{KurisuFukamiKoike2025} establish neural learning bounds for
dependent data. The following sufficient bound adds the simultaneous
control needed for our chronological fits.

\begin{corollary}[Simultaneous neural learning and inference]
\label{cor:nn-rates}
Under Appendix Assumption~\ref{ass:nn-technical}, set
\begin{equation}\label{eq:nn-rate}
 r_{N,n}:=d_n+
 \sqrt{\frac{v_n\{H_n+\log(enK_n)\}}{T_n}}
 +\sqrt{\varepsilon_n}+T_n^{-1}.
\end{equation}
If $r_{N,n}\to0$, then
$\max_{j,k}\norm{\widehat p_{jk}-p_{j,0}}_2\le Cr_{N,n}$ with
probability tending to one. The localization bounds hold with
$B_n=Cr_{N,n}^2$ and $\rho_n=Cr_{N,n}$, and the oracle score is bounded.
Under the covariance, chronology and bandwidth conditions of
Corollary~\ref{cor:wald}, with $n\beta_n(b_n)\to0$,
feasible inference follows if
\[
 \sqrt n r_{N,n}^2\to0,\qquad
 \sqrt{b_n+1}r_{N,n}\to0,\qquad
 \sqrt{L_n}r_{N,n}\to0.
\]
\end{corollary}

Bounded-parameter networks with Lipschitz activations on compact
predictor support admit finite
parameter-grid covers; Supplementary Section~\external{sec:nn-proofs}
gives the construction. Training-only validation of weight decay
$\widehat\lambda\le\kappa_{\max}/T_n$ is also compatible with the
bound. Under the parameter restrictions and global penalized
optimization gap $\varepsilon_{\mathrm{pen},n}$ in Supplementary
Proposition~\external{prop:nn-validation}, a final refit on all $T_n$
observations permits
$\varepsilon_n=2\varepsilon_{\mathrm{pen},n}
+9\kappa_{\max}(N_{\mathrm{par},n}-1)/T_n$.

\section{Simulation evidence}\label{sec:simulations}

The experiments address three questions: how much does correction reduce
size distortion when training and evaluation objectives differ; how does
the improvement vary with training sample size, model complexity and
refitting frequency; and what are the consequences for local power?
We study the state-weighted lasso comparison in
Example~\ref{ex:state-evaluation} and nonlinear probability forecasts
under the cost-weighted loss in Example~\ref{ex:probability-evaluation}.
Both experiments test equality of population evaluation risks. Write
$n$ for the evaluation sample size and $T$ for the training window length.

For raw and orthogonal scores, we compare Wald and score-bootstrap tests
at the two-sided nominal 5\% level, with Bartlett long-run variance
estimates. The four procedures are labeled Raw, Bootstrap Raw,
Orthogonal and Bootstrap Orthogonal.
The centered circular block bootstrap recomputes studentized statistics
while holding fitted objects fixed,
using 499 resamples for lasso and 999 for neural forecasts. It examines
whether resampling evaluation scores alone improves calibration.
Comparable configurations share simulated paths. Figures report
pointwise 95\% Wilson Monte Carlo intervals; the supplement gives the
designs and paired comparisons.

\subsection{Lasso complexity and training information}

The forecasting dictionaries omit a state--predictor interaction, so
the population least-squares conditions do not eliminate the derivative
of the state-weighted evaluation risk. The two nonnested dictionaries
have $p\in\{32,128,256\}$ coefficients; added coordinates are irrelevant,
so population targets and sparsity remain
fixed. Chronological cross-validation (CV) selects separate forecast and
correction penalties, and both regressions use the same eligible
training window. The stationary predictor and state vector is retained
with probability 0.5 and otherwise redrawn independently. A gap separates
training from evaluation. Supplementary Section~\external{app:sim-lasso}
specifies the construction and distinguishes CV tuning from the
sufficient penalty calibration in Corollary~\ref{cor:lasso}.

\begin{figure}[!htbp]
\centering
\includegraphics[width=\textwidth]{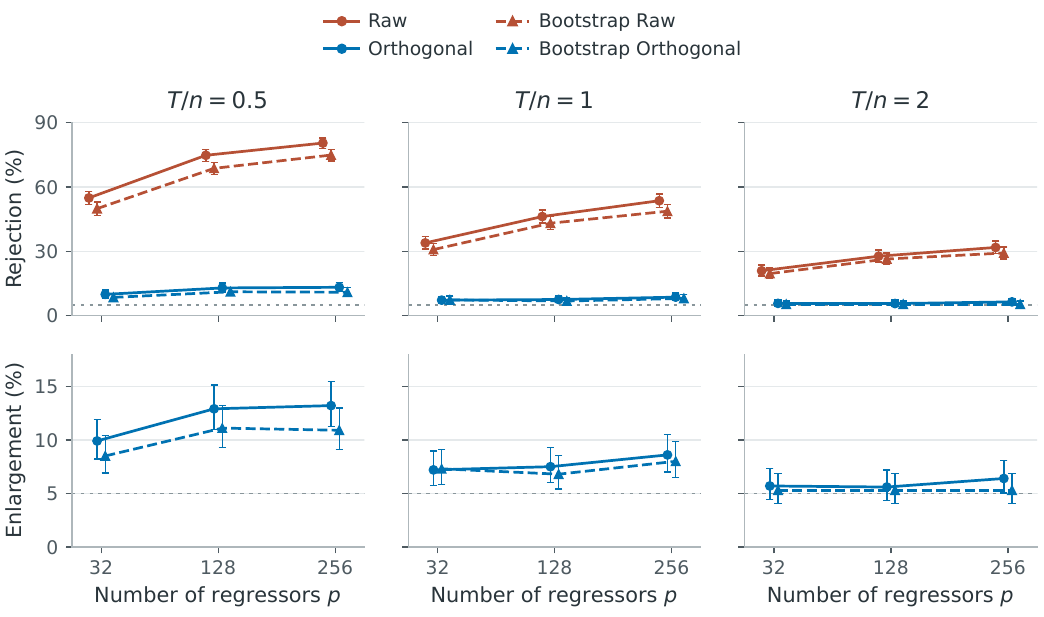}
\caption{Lasso size across dictionary sizes and training histories.}
\label{fig:simulation-size}
\begin{singlespace}\footnotesize\raggedright
\noindent Notes: $n=1{,}000$, five chronological fits, CV tuning,
1,000 replications per point. Rejection percentages at nominal 5\%.
The lower row enlarges the region near nominal size. Bootstrap
comparisons hold fitted objects fixed.
\end{singlespace}
\end{figure}

Correction substantially reduces size distortion across dictionary sizes,
and longer training brings rejection closer to nominal size
(Figure~\ref{fig:simulation-size}). At $n=1{,}000$ with five fits and
$T=n$, raw rejection rises from 33.9\% to 53.6\%
as $p$ increases, compared with 7.2--8.6\% after correction. Bootstrap
raw rejection remains 30.8--48.7\%. With $T=2n$, corrected rejection
is 5.6--6.4\%, while raw rejection is 20.9--31.8\%. The ratio
$p/T$ indexes these comparisons; the theoretical bounds depend on
sparsity and $\log p$.

Short training windows leave appreciable overrejection: at $T=0.5n$,
corrected rejection is 9.9--13.2\%, against raw rates of 54.9--80.5\%.
The diagnostics link these differences to the population mean error of
the score with fitted objects held fixed. At $T=n,p=32$, this error,
scaled by $\sqrt n$ and averaged across replications, falls from 1.69
to 0.17. Coverage of nominal 95\% Wald intervals rises from 66.1\% to 92.8\%,
with nearly unchanged mean length. Further diagnostics in Supplementary
Section~\external{app:sim-replacement} reinforce this interpretation:
orthogonal correction removes a substantial drift component, while
finite-sample inference remains sensitive to errors in estimating the
forecasts and correction functions.

\subsection{Overlapping windows and repeated refitting}

\begin{figure}[!htbp]
\centering
\includegraphics[width=\textwidth]{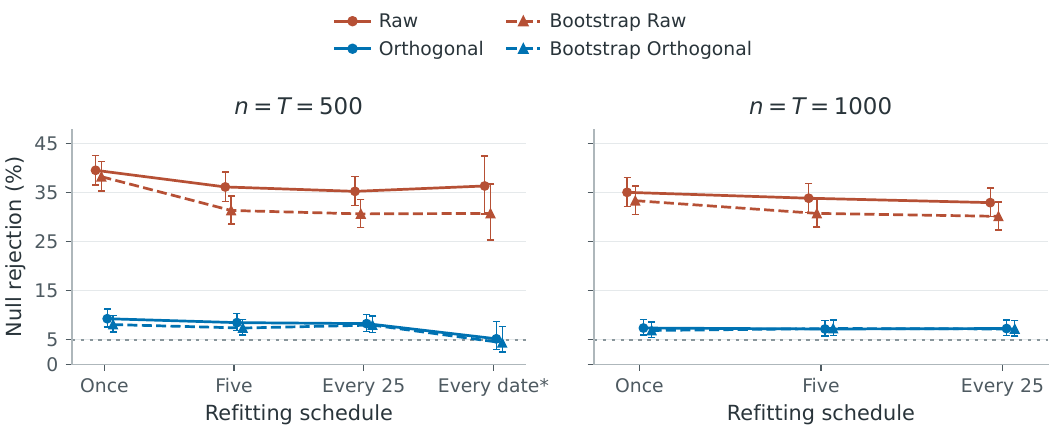}
\caption{Lasso size under repeated chronological refitting.}
\label{fig:simulation-refits}
\begin{singlespace}\footnotesize\raggedright
\noindent Notes: $T=n$, $p=32$, CV tuning. Standard cells use 1,000
replications; every-date refitting uses 250 at $n=500$ and is unavailable
at $n=1{,}000$. Methods and intervals follow Figure~\ref{fig:simulation-size}.
\end{singlespace}
\end{figure}

The size improvement persists under repeated estimation on overlapping
windows. Figure~\ref{fig:simulation-refits} compares one fit, five fits
and refitting every 25 evaluation dates at $T=n,p=32$. At $n=1{,}000$,
orthogonal rejection stays at 7.2--7.4\%, compared with raw rates of
33.0--35.1\%; at $n=500$, the respective ranges are 8.3--9.3\% and
35.3--39.6\%. The additional every-date experiment at $n=T=500$ gives corrected
and raw rejection rates of 5.2\% and 36.4\%. Its 250 replications do not 
permit precise comparisons across schedules, but the reduction in
distortion extends to refitting at every evaluation date.

The supplement compares power at $\theta=\delta/\sqrt n$ after calibrating
each test to 5\% rejection under its matching empirical null. At
$n=T=1{,}000,p=32$,
five fits and $\delta=-0.7$, size-adjusted power rises from 1.2\%
to 8.4\%. Correction restores sensitivity against an alternative obscured
by estimation drift, although power remains modest. The power gain is not uniform
across signs. Supplementary Sections~\external{app:sim-training}
and~\external{app:sim-power} report further training and refitting comparisons.

\subsection{Nonlinear ReLU probability forecasts}

The probability experiment combines nonlinear forecasts with an analytic
correction. Each rule observes a different set of five predictors entering the true
probability through polynomial terms and interactions. Transforming
Gaussian AR(1) coordinates with coefficient 0.5 gives uniform predictors.
Networks have two hidden ReLU layers of width eight, an affine skip
connection and bounded output. Training uses squared loss with a quadratic
penalty; evaluation uses $(1+4Y)(Y-p)^2$, with the analytic correction in
\eqref{eq:nn-score}. At each of five chronological fits, training-only
validation selects the penalty and number of training iterations.
Supplementary Section~\external{app:sim-neural} gives the DGP and
distinguishes this empirical penalty choice from the sufficient
calibration accompanying Corollary~\ref{cor:nn-rates}.

\begin{figure}[!htbp]
\centering
\includegraphics[width=\textwidth]{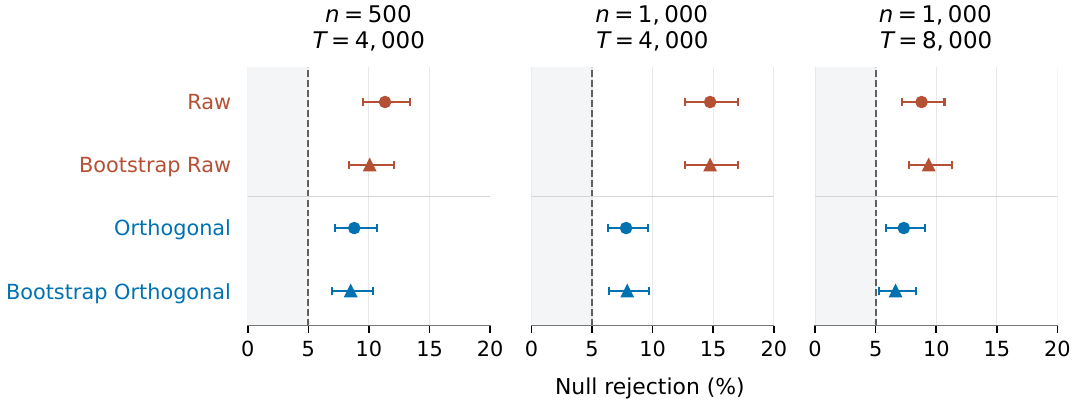}
\caption{Size for nonlinear ReLU probability forecasts.}
\label{fig:simulation-neural}
\begin{singlespace}\footnotesize\raggedright
\noindent Notes: Five chronological fits, 1,024 replications per cell.
All panels use the same scale. Methods and intervals follow
Figure~\ref{fig:simulation-size}. Pilot draws are excluded.
\end{singlespace}
\end{figure}

Correction also improves size for the nonlinear forecasts, although
overrejection remains (Figure~\ref{fig:simulation-neural}). At
$n=1{,}000,T=4{,}000$, raw and bootstrap raw rejection are both 14.7\%,
compared with 7.8\% for the orthogonal test and 7.9\% for its bootstrap
version.
Improvements are smaller at $n=500$ and with longer training; in the
latter case, the paired raw--orthogonal difference is not statistically
significant at 5\%. Bootstrapping the corrected score provides no
uniform additional improvement.

The improvement at $n=1{,}000,T=4{,}000$ persists across the covariance
bandwidth and bootstrap block lengths examined in the supplement.
Diagnostics show reduced absolute population
drift and improved agreement between estimated standard errors and
sampling dispersion. Power can nevertheless fall: at $\delta=-0.6$,
size-adjusted power is 17.8\% after correction versus 30.5\% for the raw test.
The supplementary diagnostics show greater sampling dispersion after
correction and residual bias that attenuates this negative alternative;
the power analysis reports all eight signed alternatives. The neural experiment
thus extends the size improvement to nonlinear forecasts while showing
that better size need not imply greater size-adjusted power.
\FloatBarrier

\section{Conclusion}

Population predictive risk inference can remain feasible when machine
learning forecasts do not minimize the evaluation objective. Standard
locally robust corrections supply the required mean-error control;
our chronological replacement and covariance bounds establish Gaussian
inference with overlapping training windows and repeated fitting.
Uniform coverage additionally requires uniform learning conditions and
a uniform positive lower bound on the smallest eigenvalue of the
corrected-score long-run covariance matrix. The lasso and neural implementations
make the required forecast and correction accuracy explicit.
Simulations show substantial size improvements for lasso and nonlinear
ReLU probability forecasts, beyond those obtained by bootstrapping raw
scores. The diagnostics reveal remaining learning
effects and a possible power cost: orthogonality removes first-order
sensitivity, but does not ensure accurate finite-sample inference or
uniform power gains.

\section*{Declarations}

\paragraph{Funding and computing resources.}
The author received no financial funding for this research. Computations
were performed using the Dutch national supercomputer Snellius, operated by SURF.

\paragraph{Competing interests.}
The author declares no competing interests.

\paragraph{Data and code availability.}
All simulation data are generated by the accompanying replication code,
available at \url{https://github.com/a91quaini/replicate-oos-ml-test}.

\paragraph{Declaration of generative AI and AI-assisted technologies in the manuscript preparation process.}
During the preparation of this work, the author used OpenAI's ChatGPT to assist 
with coding, language editing, and reviewing and refining mathematical proofs. 
The author reviewed and edited the content as needed and takes full responsibility 
for the content of the manuscript.
\appendix
\renewcommand{\thesection}{\Alph{section}}
\setcounter{assumption}{0}
\renewcommand{\theassumption}{\thesection\arabic{assumption}}
\renewcommand{\theHassumption}{appendix.\thesection.\arabic{assumption}}
\setcounter{equation}{0}
\renewcommand{\theequation}{\thesection.\arabic{equation}}
\renewcommand{\theHequation}{appendix.\thesection.\arabic{equation}}
\section{Technical conditions}
\label{sec:technical-conditions}

This appendix collects the technical conditions used by the main results.
Constants are uniform in the sample index in triangular-array applications.

\subsection{Population maps and local remainders}\label{sec:technical-population}

Prediction, loss and score maps are jointly measurable in their data
and parameter arguments; losses and scores used in the estimators are
finite on those arguments. All displayed expectations are assumed finite.
Fitted nuisances are jointly measurable in training observations and fitting
randomness, and all membership events used in localization are measurable;
the deterministic neighborhoods in Assumption~\ref{ass:learning}
are measurable. Derivative norms are operator norms unless otherwise stated.

\begin{assumption}[Population differentiability]\label{ass:population}
Let $\mathcal U_j$ be a convex open neighborhood of $\eta_{j,0}$ in $\Hh_j$. On $\mathcal U_j$, $Q_j$ is twice continuously Fr\'echet differentiable,
$DQ_j(\eta_{j,0})=0$, and each $R_{j,r}$ is continuously Fr\'echet
differentiable. The training directional scores have jointly
measurable versions $s_{j,t}(\eta_j)[v_j]$, are integrable for fixed
arguments, and satisfy the score identity
\eqref{eq:scoreidentity}.
For each fixed $\eta_j$, the score is linear in its direction argument,
with the linearity identities holding almost surely for each fixed
choice of directions.
\end{assumption}

\begin{assumption}[Local smoothness and score moments]\label{ass:local-technical}
\textup{(i)} On the neighborhoods of Assumption~\ref{ass:population}, require
$\sup_{\eta_j\in\mathcal U_j}\norm{D^2Q_j(\eta_j)}\le C_Q$.
For
\[
 T_{j,r}(\eta_j):=\epsilon_jR_{j,r}(\eta_j)
                         -DQ_j(\eta_j)[\alpha_{j,r,0}],
\]
suppose $T_{j,r}$ is twice continuously Fr\'echet differentiable with
Hessian norm bounded by $C_T$ on $\mathcal U_j$.
Alternatively, the same local mean bound follows if each $R_{j,r}$
is twice continuously Fr\'echet differentiable,
$\sup\norm{D^2R_{j,r}}\le C_R$, and
\[
 \norm{D^2Q_j(\eta_j)-D^2Q_j(\widetilde\eta_j)}
                  \le C_H\norm{\eta_j-\widetilde\eta_j},
\]
with bounded $\norm{\alpha_{j,r,0}}$.

\textup{(ii)} Writing $u_j=\eta_j-\eta_{j,0}$, assume the following
bounds uniformly on these neighborhoods and, in
\eqref{eq:scorelocal3}, for every $v_j\in\Hh_j$:
\begin{align}
 \norm{g_t(\eta)-g_t(\eta_0)}_2&\le C_g\sum_j\norm{u_j},\label{eq:scorelocal1}\\
 \norm{s_{j,t}(\eta_j)[\alpha_{j,r,0}]
              -s_{j,t}(\eta_{j,0})[\alpha_{j,r,0}]}_2
              &\le C_s\norm{u_j},\label{eq:scorelocal2}\\
 \norm{s_{j,t}(\eta_j)[v_j]}_2&\le C_s\norm{v_j}.
              \label{eq:scorelocal3}
\end{align}
\end{assumption}

The $L^2$ score conditions in Assumption~\ref{ass:local-technical}
concern the actual observation-level maps.
They do not require point evaluation to be a continuous functional on a
space of prediction functions. They may be replaced by any directly
verified local $L^2$ bound when applying the inference results.

\subsection{Observation array and score moments}\label{sec:technical-array}

For sigma fields $\mathcal B,\mathcal C$, define absolute regularity by
\[
 \beta(\mathcal B,\mathcal C)
 :=\frac12\sup_{(B_i),(C_j)}
       \sum_{i,j}|\pr(B_i\cap C_j)-\pr(B_i)\pr(C_j)|,
\]
where the supremum ranges over finite partitions $(B_i)$ and $(C_j)$
of the sample space with $B_i\in\mathcal B$ and $C_j\in\mathcal C$.
Put
$\beta_n(h):=\sup_t\beta\{\sigma(Z_{s,n}:s\le t),
\sigma(Z_{s,n}:s\ge t+h)\}$.

\begin{assumption}[Observation array and oracle moments]\label{ass:array}
For each $n$, $(Z_{t,n}:t\in\mathbb Z)$ is strictly stationary on a
standard Borel space and
$\beta_n(h)\le C_\beta e^{-c_\beta h}$ for fixed positive constants.
For some $\delta>0$,
\begin{equation}\label{eq:oraclemoment}
               \sup_n\E\{\norm{\xi_{0,n}}^{4+\delta}\}\le C.
\end{equation}
\end{assumption}

\subsection{Regularity for the forecasting procedures}
\label{sec:learner-regularity}

These conditions supply the primitive learning bounds for the procedures
in Section~\ref{sec:learner-summary}. They do not change the population
training or evaluation targets. All constants are uniform in the array
index and training window.

\begin{assumption}[Mean and projection corrections]
\label{ass:mean-technical}
The outcome satisfies $\E Y^2<\infty$, and $q$ is bounded. For every
outcome $y$, the derivative $z\mapsto\dot\ell(y,z)$ is Lipschitz with
a common constant $L$. At the population forecast $f_{j,0}$,
$q_r\ell(Y,f_{j,0})\in L^1(P)$ and
$q_r\dot\ell(Y,f_{j,0})\in L^2(P)$ for every $j,r$.
\end{assumption}

\begin{assumption}[Sparse forecasts and generated corrections]
\label{ass:lasso-technical}
The observations are strictly stationary and geometrically beta-mixing,
with mixing constants uniform in $n$. The instruments are bounded, and
$Y$ and the deterministic predictor dictionaries $X_j\in\R^{p_n}$ are
uniformly sub-Gaussian:
\[
 \sup_{j,n}\sup_{\|v\|=1}\|v^{\T}X_j\|_{\psi_2}<\infty,
 \qquad \sup_n\|Y\|_{\psi_2}<\infty,
\]
where $\|U\|_{\psi_2}=\inf\{c>0:\E\exp(|U|^2/c^2)\le2\}$.
The population Gram matrices satisfy
$\lambda_{\min}(\Sigma_j)\ge c>0$.
The projection and representer coefficients have at most $s_{f,n}$ and
$s_{a,n}$ nonzero entries, respectively. In
\eqref{eq:lasso-fit}--\eqref{eq:lasso-representer}, use measurable
minimizers, penalize every coefficient, and take
$\lambda_n=C_\lambda r_n$, where $r_n$ is defined in
\eqref{eq:lasso-rate} and $C_\lambda$ is sufficiently large.
The sparsity and concentration scales satisfy
\((1+s_{f,n}+s_{a,n})r_n\longrightarrow0\).
\end{assumption}

\begin{assumption}[Neural classes and empirical optimization]
\label{ass:nn-technical}
The binary-outcome observation array is strictly stationary and uniformly
geometrically beta-mixing. For each rule, $\mathcal P_{j,n}$ is a
deterministic measurable neural class with outputs in $[0,1]$.
Define its population approximation error and logarithmic covering number by
\[
 d_n=\max_j\inf_{p\in\mathcal P_{j,n}}\norm{p-p_{j,0}}_2,
 \qquad
 H_n=\max_j\log N(T_n^{-2},\mathcal P_{j,n},\norm\cdot_\infty)<\infty,
\]
where $N$ denotes the covering number on the predictor support.
Each fitted forecast belongs to its declared class and, simultaneously
over $j,k$, satisfies
\begin{equation}\label{eq:nn-optimization}
 \E_k(Y-\widehat p_{jk})^2
 \le\inf_{p\in\mathcal P_{j,n}}\E_k(Y-p)^2+\varepsilon_n
\end{equation}
with probability tending to one. The deterministic optimization tolerance
is $\varepsilon_n\ge0$. The dependence block length is
$v_n=\lceil C_v\log(enK_nT_n)\rceil$, with a sufficiently large fixed $C_v$.
\end{assumption}

Condition~\eqref{eq:nn-optimization} is an objective-gap requirement
relative to the global empirical infimum. Supplementary Section~\external{sec:nn-proofs}
gives the covering
argument and a sufficient bound for training-only selection of weight decay.
\bibliographystyle{plainnat}
\bibliography{references}
\end{document}


\title{Supplement to ``Orthogonal inference for machine learning forecasts''}
\author{Alberto Quaini\thanks{Corresponding author.}\\ Erasmus School of Economics, Erasmus University Rotterdam\\ P.O. Box 1738, 3000 DR Rotterdam, The Netherlands\\ \texttt{quaini@ese.eur.nl}}
\date{}
\maketitle
Numbered results without an S prefix refer to the main paper.
The main paper's notation and Appendix~\external{sec:technical-conditions}
apply throughout. Section~\ref{sec:supp-population-proofs} proves the
population results. Section~\ref{sec:supp-dependence} gives the probability
bounds used in Section~\ref{sec:supp-inference-proofs}, which proves the
inference results and the raw-score and state-normalization comparisons.
Section~\ref{sec:supp-learning} supplies the correction calculations and
learning proofs for the two forecasting implementations.
Section~\ref{sec:degeneracy} illustrates the covariance restriction, and
Section~\ref{app:simulations} reports the full simulation designs and
supporting evidence.

\section{Proofs of the population results}\label{sec:supp-population-proofs}

\begin{proof}[Proof of Theorem~\external{thm:population}]
By \eqexternal{eq:scoreidentity},
\[
 \Psi_r(\theta,\nu)
 =M_r(\eta)-\sum_jDQ_j(\eta_j)[\alpha_{j,r}]-\theta_r.
\]
The training first-order conditions and $M(\eta_0)=\theta_0$ imply
$\Psi_r(\theta_0,\nu_0)=0$. The map
$(\eta_j,\alpha_{j,r})\mapsto DQ_j(\eta_j)[\alpha_{j,r}]$ is jointly
Fr\'echet differentiable, since $DQ_j$ is continuously Fr\'echet
differentiable and evaluation of a bounded functional is continuous
bilinear. Its derivative at the target is
$A_{j,0}(u_j,\alpha_{j,r,0})+DQ_j(\eta_{j,0})[b_{j,r}]$.
Therefore
\[
 D_\nu\Psi_r(\theta_0,\nu_0)[u,b]
 =\sum_j\{G_{j,r}[u_j]
       -A_{j,0}(u_j,\alpha_{j,r,0})
       -DQ_j(\eta_{j,0})[b_{j,r}]\}=0.
\]

\end{proof}

\begin{proof}[Proof of Proposition~\external{prop:bounds}]
Write $u_j:=\eta_j-\eta_{j,0}$ and $b_{j,r}:=\alpha_{j,r}-\alpha_{j,r,0}$.
Put $T_{j,r}(\eta_j):=\epsilon_jR_{j,r}(\eta_j)
-DQ_j(\eta_j)[\alpha_{j,r,0}]$, as in Appendix
Assumption~\external{ass:local-technical}. Linearity in the representer
and the training first-order conditions give the exact decomposition
\[
 \E\{\phi_{t,r}(\nu)\}-\theta_{0,r}
 =\sum_j\{T_{j,r}(\eta_j)-T_{j,r}(\eta_{j,0})
                   -DQ_j(\eta_j)[b_{j,r}]\}.
\]
The representer identity gives $DT_{j,r}(\eta_{j,0})=0$.
Taylor's integral formula therefore bounds
$|T_{j,r}(\eta_j)-T_{j,r}(\eta_{j,0})|$ by
$C_T\norm{u_j}^2/2$. The training first-order condition and integration along the segment
in $\mathcal U_j$ give
\[
 |DQ_j(\eta_j)[b_{j,r}]|
 =\left|\int_0^1D^2Q_j(\eta_{j,0}+tu_j)[u_j,b_{j,r}]\,dt\right|
 \le C_Q\norm{u_j}\norm{b_{j,r}}.
\]
Sum over $j$ and take the Euclidean norm across the fixed number of
components to obtain
\begin{equation}\label{eq:biasbound-full}
 \norm{\E\{\phi_t(\nu)\}-\theta_0}
 \le C\sum_j\left\{\norm{u_j}^2+
       \norm{u_j}\left(\sum_r\norm{b_{j,r}}^2\right)^{1/2}\right\},
\end{equation}
with $C$ depending only on $d,C_Q,C_T$.

Under the alternative smoothness conditions in Appendix
Assumption~\external{ass:local-technical}, the Taylor remainder in
$R_{j,r}$ is at most $C_R\norm{u_j}^2/2$, while
\[
 |DQ_j(\eta_j)[\alpha_{j,r,0}]
       -A_{j,0}(u_j,\alpha_{j,r,0})|
 \le\int_0^1 C_Ht\norm{u_j}^2\norm{\alpha_{j,r,0}}\,dt.
\]
The linear terms cancel by \eqexternal{eq:riesz}; the product term is
bounded as before. The bound \eqref{eq:biasbound-full} then holds with
$C_T$ replaced by $C_R+C_H\max_{j,r}\norm{\alpha_{j,r,0}}$.

For fixed arguments, linearity of the training score gives
\begin{align*}
 \phi_{t,r}(\nu)-\phi_{t,r}(\nu_0)
 ={}&g_{t,r}(\eta)-g_{t,r}(\eta_0)\\
 &-\sum_j\{s_{j,t}(\eta_j)[\alpha_{j,r,0}]
                     -s_{j,t}(\eta_{j,0})[\alpha_{j,r,0}]\}
       -\sum_js_{j,t}(\eta_j)[b_{j,r}].
\end{align*}
Minkowski's inequality and \eqexternal{eq:scorelocal1}--\eqexternal{eq:scorelocal3}
give
\[
 \norm{\phi_t(\nu)-\phi_t(\nu_0)}_2
 \le C\sum_j\left\{\norm{u_j}
          +\left(\sum_r\norm{b_{j,r}}^2\right)^{1/2}\right\}.
\]
Substituting the neighborhood radii gives \eqexternal{eq:biasbound}
and the scales in \eqexternal{eq:rates}. All population expectations hold their nuisance
arguments fixed.
\end{proof}

\section{Supporting theory for chronological inference}\label{sec:supp-dependence}
We use Berbee coupling in the form stated in
\citet[Section 3, pp.~406--407]{DoukhanMassartRio1995}:
on an enlarged probability space, a random element has a copy with
the same law, independent of a specified sigma field, with mismatch
probability equal to their absolute regularity coefficient.
Successive coupling gives independent copies of separated blocks
with total mismatch probability bounded by the sum of the coefficients.
The observation-space and measurability conditions in
Appendix~\external{sec:technical-conditions} permit these constructions.
We also use the conditional total variation formulation of absolute regularity.

The mixing covariance inequality recalled in
\citet[equation~(1.0)]{Rio1993} yields the following uniform bounds for
the Bartlett estimator of \citet{NeweyWest1987}. We retain the argument
to make the dependence on the bandwidth and the uniformity across rows explicit.

\begin{lemma}[Uniform covariance and HAC bounds]\label{lem:moments}
Let $V_{t,n}=v_n(Z_{t,n})$, with measurable $v_n$, be a centered
stationary vector array of fixed dimension satisfying
Assumption~\external{ass:array} with $V$ in place
of $\xi$. Uniformly in $n$, its covariance matrices have an
exponential envelope, and for any four coordinates $a,b,c,e$,
\begin{equation}\label{eq:covexp}
 \norm{\cov(V_{0,n},V_{h,n})}\le C e^{-c_0|h|},\qquad
 \sum_{r,s,t\in\mathbb Z}
 |\cum(V_{0,n}^{(a)},V_{r,n}^{(b)},V_{s,n}^{(c)},V_{t,n}^{(e)})|
                         \le C.
\end{equation}
With $\Omega_{V,n}:=\sum_h\cov(V_{0,n},V_{h,n})$,
\begin{equation}\label{eq:hacrate}
 \norm{\mathcal B_n(V)-\Omega_{V,n}}
       =\Op\!\left(\sqrt{L_n/n}\right)+O(L_n^{-1}+n^{-1})
\end{equation}
whenever $L_n\to\infty$, $L_n<n$.
\end{lemma}

\begin{proof}[Proof of Lemma~\ref{lem:moments}]
For random variables separated by $h$ dates, the mixing covariance
inequality gives
\[
 |\cov(A,B)|\le C\norm A_p\norm B_q
                \beta_n(h)^{1-1/p-1/q},\qquad p^{-1}+q^{-1}<1.
\]
This follows from \citet[equation~(1.0)]{Rio1993}, since strong mixing
coefficients are bounded by absolute regularity coefficients.
Applying it to coordinates proves the first inequality in
\eqref{eq:covexp}. For centered
scalar variables, the fourth cumulant is their fourth product
expectation minus the three products of pairwise expectations.
Order four time indices and denote their largest adjacent gap by $g$.
If that gap divides the indices two and two, the cumulant equals
the covariance of the two products minus the two remaining
covariance products. If it divides them one and three, the fourth
product expectation is the covariance of the singleton and triple
product, since the singleton is centered. In the remaining
covariance products at least one pair crosses the largest gap.
For products of $r$ and $4-r$ coordinates, H\"older's inequality
gives moment exponents $(4+\delta)/r$ and $(4+\delta)/(4-r)$.
The covariance inequality bounds every relevant term by
$Ce^{-c_1g}$. Repeated indices satisfy the same bound with a larger
constant. Fixing the position of time zero and summing over ordered
gaps proves the second inequality in \eqref{eq:covexp}, because
$\sum_{g_1,g_2,g_3\ge0}e^{-c_1\max(g_1,g_2,g_3)}<\infty$.

Suppress the row index and put
$w_h=(1-|h|/(L_n+1))\ind\{|h|\le L_n\}$. Before sample centering,
the Bartlett estimator is the quadratic form
\[
 B_n^\circ:=\frac1n\sum_{t,u=1}^n w_{t-u}V_tV_u^{\T}.
\]
For coordinates $a,b$, expand
$\var\{(B_n^\circ)_{ab}\}$ as a sum over $t,u,s,v$.
The covariance of $V_t^{(a)}V_u^{(b)}$ and
$V_s^{(a)}V_v^{(b)}$ is a fourth cumulant plus two products
of cross covariances. Since $|w_h|\le1$, stationarity and the
cumulant bound in \eqref{eq:covexp} bound the absolute cumulant
contribution by
\[
 \frac1{n^2}\sum_{t,u,s,v=1}^n
 |\cum(V_t^{(a)},V_u^{(b)},V_s^{(a)},V_v^{(b)})|
 \le \frac Cn.
\]
Write $\Gamma_{ab}(h):=\E(V_0^{(a)}V_h^{(b)})$.
For the covariance product
$\Gamma_{aa}(s-t)\Gamma_{bb}(v-u)$, fix the two displacements
$p=s-t$ and $q=v-u$. At most $n(2L_n+1)$ pairs $(t,u)$
have $w_{t-u}\ne0$, and the other weight is bounded by one.
Its absolute contribution is therefore at most
\[
 \frac{2L_n+1}{n}
       \sum_{p,q\in\mathbb Z}|\Gamma_{aa}(p)\Gamma_{bb}(q)|
 \le C L_n/n.
\]
Fixing $p=v-t$ and $q=s-u$ gives the same bound for the other
covariance product. Thus, in fixed dimension,
$\norm{B_n^\circ-\E B_n^\circ}=\Op(\sqrt{L_n/n})$.
Its expectation weights lag $h$ by $w_h(1-|h|/n)$.
Exponential covariance summability, including
$\sum_h|h|\norm{\cov(V_0,V_h)}<\infty$, bounds the Bartlett
and truncation bias by $C/L_n$ and the endpoint bias by $C/n$.

To control sample centering, put
\[
 d_t:=\sum_{u=1}^n w_{t-u},\qquad
 A_n:=\frac1n\sum_{t=1}^n d_tV_t,\qquad
 c_n:=\frac1n\sum_{t=1}^n d_t.
\]
The weights are nonnegative and $d_t\le L_n+1$, so the covariance
envelope gives $\E\norm{A_n}^2\le C L_n^2/n$ and
$\bar V_n=\Op(n^{-1/2})$. Expanding the centered quadratic form gives
\[
 \mathcal B_n(V)-B_n^\circ
 =-A_n\bar V_n^{\T}-\bar V_nA_n^{\T}
                        +c_n\bar V_n\bar V_n^{\T}
 =\Op(L_n/n),
\]
where $c_n\le L_n+1$. Since $L_n<n$, this term is absorbed by
$\Op(\sqrt{L_n/n})$, proving \eqref{eq:hacrate}.
\end{proof}

\citet[Theorem~5]{BallinariWehrli2025} establish feasible long-run
variance estimation for cross-fitted orthogonal impulse-response scores.
For our chronological scores, the positive-semidefinite Bartlett
construction of \citet[Theorem~1]{NeweyWest1987} gives the following
deterministic covariance comparison.

\begin{lemma}[Deterministic covariance comparison]\label{lem:haccompare}
For vector sequences $x,y$, write
$\norm x_n:=(n^{-1}\sum_t\norm{x_t}^2)^{1/2}$ and put $e=x-y$. Then
\begin{align}
 \norm{\mathcal B_n(x)-\mathcal B_n(y)}
 &\le 2\{\operatorname{tr}\mathcal B_n(y)
                 \operatorname{tr}\mathcal B_n(e)\}^{1/2}
                     +\operatorname{tr}\mathcal B_n(e)\nonumber\\
 &\le 2\{(L_n+1)\operatorname{tr}\mathcal B_n(y)\}^{1/2}
                  \norm e_n+(L_n+1)\norm e_n^2.
                  \label{eq:haccompare-sharp}
\end{align}
For any deterministic matrix $A$,
$\mathcal B_n(Ax)=A\mathcal B_n(x)A^{\T}$.
\end{lemma}

\begin{proof}[Proof of Lemma~\ref{lem:haccompare}]
First center the sequences. Centering is an orthogonal projection
on the Euclidean sample space and does not increase either
$\norm y_n$ or the difference norm. Put $e=x-y$ after centering.
For a centered sequence $z$, set $z_t=0$ outside the sample and, for
$s\in\mathbb Z$, put
\[
 U_s(z):=\{n(L_n+1)\}^{-1/2}\sum_{t=s}^{s+L_n}z_t.
\]
There are $(L_n+1-|t-u|)_+$ windows containing both dates $t,u$.
Counting these windows gives the Bartlett block-sum identity
$\mathcal B_n(z)=\sum_{s\in\mathbb Z}U_s(z)U_s(z)^{\T}$;
only finitely many terms are nonzero. This also proves positive
semidefiniteness.
Expanding at $x=y+e$ and applying Cauchy--Schwarz to the two
cross terms bounds the covariance difference by
\[
 2\left\{\sum_s\norm{U_s(y)}^2
                 \sum_s\norm{U_s(e)}^2\right\}^{1/2}
                       +\sum_s\norm{U_s(e)}^2.
\]
This is the first inequality in \eqref{eq:haccompare-sharp}.
Each observation enters $L_n+1$ windows, so another
Cauchy--Schwarz inequality gives
\[
 \operatorname{tr}\mathcal B_n(e)
 \le n^{-1}\sum_s\sum_{t=s}^{s+L_n}\norm{e_t}^2
 =(L_n+1)\norm e_n^2.
\]
Centering is contractive, so the same bound holds with the
original difference norm. This proves the second inequality.
The transformation identity follows directly from
\eqexternal{eq:bartlett}.
\end{proof}

\subsection{Chronological replacement}

\begin{lemma}[Chronological score replacement]\label{lem:replacement}
Under Assumptions~\external{ass:chronology}--\external{ass:learning} and the mixing part of
Assumption~\external{ass:array}, put
\[
 m_n(\nu):=\E\{\phi_{t,n}(\nu)-\phi_{t,n}(\nu_{0n})\},
\]
where the expectation holds $\nu$ fixed. If
\[
 n\beta_n(b_n)\to0,\qquad
 \sqrt{b_n+1}\,\rho_n\to0,
\]
then
\begin{equation}\label{eq:centeredreplacement}
 \frac1{\sqrt n}\sum_{t=1}^n
 \{\phi_t(\widehat\nu_{k(t)})-\phi_t(\nu_{0n})
                        -m_n(\widehat\nu_{k(t)})\}=\op(1).
\end{equation}
If also $\sqrt n B_n\to0$, then
\begin{equation}\label{eq:meanreplacement}
 \frac1{\sqrt n}\sum_{t=1}^n
 \{\phi_t(\widehat\nu_{k(t)})-\phi_t(\nu_{0n})\}=\op(1).
\end{equation}
For covariance replacement, localization, the local $L^2$ bound and
$n\beta_n(b_n)\to0$ alone imply
\begin{equation}\label{eq:rmsreplacement}
 \left[n^{-1}\sum_{t=1}^n
 \norm{\phi_t(\widehat\nu_{k(t)})-\phi_t(\nu_{0n})}^2\right]^{1/2}
                     =\Op(\rho_n).
\end{equation}
Thus the centered replacement uses score accuracy and temporal
separation, while the uncentered replacement also uses the mean bound.
\end{lemma}

\begin{proof}[Proof of Lemma~\ref{lem:replacement}]
\emph{Localization and conditional laws.}
Replace each fit by $\nu_{0n}$ whenever it lies outside $\mathcal N_n$.
The resulting fits retain training measurability and agree simultaneously
with the original fits on the localization event, whose probability tends
to one. We first establish the bounds for these localized fits. Put
$d_n=b_n+1$, $r_n=\rho_n$, $\varepsilon_n=\beta_n(b_n)$ and
\[
 D_t(z):=\{\phi_n(z;\widehat\nu_{k(t)})-\phi_n(z;\nu_{0n})\}.
\]
For $t\in I_k$, its random parameter is measurable with respect to
$\mathcal G_t:=\Ff_{t-d_n,n}$ since $t_k\le t$.
On the standard Borel observation space, let $Q_t$ be a regular
conditional law of $Z_t$ given $\mathcal G_t$. The total variation
distance $V_t=\sup_B|Q_t(B)-P_n(B)|$ is measurable: the supremum
can be taken over a countable generating algebra. Absolute
regularity gives $\E V_t\le C\varepsilon_n$. The independent
fitting randomness does not change this bound. Hence for a random,
jointly measurable, $\mathcal G_t$-measurable function $f_t$ bounded
in norm by a deterministic $M$,
\begin{equation}\label{eq:tv}
 \E\left\|\int f_t\,dQ_t-\int f_t\,dP_n\right\|
                         \le CM\varepsilon_n.
\end{equation}
This follows pathwise from the total variation bound and then by
expectation, so the function may depend on the whole training history.

\emph{Centered score error.}
Truncate $D_t$ at norm $h_n:=\sqrt{n/d_n}$:
$H_t(z):=D_t(z)\ind\{\norm{D_t(z)}\le h_n\}$ and $X_t:=H_t(Z_t)$.
The local bound gives, for every realized localized parameter,
\[
 P_n\{\norm{D_t}>h_n\}\le Cr_n^2/h_n^2,\qquad
 \left\|\int(H_t-D_t)\,dP_n\right\|\le Cr_n^2/h_n.
\]
Apply \eqref{eq:tv} to the tail indicator and take a union bound:
\[
 \pr\{X_t\ne D_t(Z_t)\text{ for some }t\}
       \le Cn\varepsilon_n+Cn r_n^2/h_n^2
       =C(n\varepsilon_n+d_nr_n^2)\to0.
\]
Set $M_t:=X_t-\E(X_t\mid\mathcal G_t)$. It is
$\Ff_{t,n}$-measurable and has conditional mean zero given
$\Ff_{t-d_n,n}$. If $s\le t-d_n$, conditional expectation gives
$\E(M_sM_t^{\T})=0$. The increments in each residue class modulo
$d_n$ are thus pairwise uncorrelated. Conditional variance
contraction and \eqref{eq:tv} applied to $\norm{H_t}^2$ yield
$\E\norm{M_t}^2\le C(r_n^2+h_n^2\varepsilon_n)$.
Cauchy--Schwarz across the residue classes, allowing empty classes
when $d_n>n$, gives
\[
 \E\left\|n^{-1/2}\sum_tM_t\right\|^2
 \le \frac{d_n}{n}\sum_t\E\norm{M_t}^2
 \le C(d_nr_n^2+n\varepsilon_n)\to0.
\]
Applying \eqref{eq:tv} to $H_t$ also gives
\[
 \E\left\|\frac1{\sqrt n}\sum_t
 \left\{\E(X_t\mid\mathcal G_t)-\int H_t\,dP_n\right\}\right\|
 \le C\sqrt n h_n\varepsilon_n
 =Cn\varepsilon_n/\sqrt{d_n}\to0.
\]
The normalized sum of marginal truncation errors is at most
$C\sqrt{d_n}r_n^2=o(1)$, since $d_nr_n^2\to0$ and $d_n\ge1$.
It follows that
\[
 n^{-1/2}\sum_t\left\{D_t(Z_t)-\int D_t\,dP_n\right\}=\op(1).
\]
Since $\int D_t\,dP_n=m_n(\widehat\nu_{k(t)})$, this is
\eqref{eq:centeredreplacement} for the localized fits.

\emph{Population mean error.}
On the localization event, the remaining marginal means satisfy
\[
 \left\|\frac1{\sqrt n}\sum_t
                      m_n(\widehat\nu_{k(t)})\right\|
                     \le C\sqrt n B_n.
\]
The additional mean restriction therefore gives
\eqref{eq:meanreplacement}.

\emph{Empirical squared error.}
For \eqref{eq:rmsreplacement}, couple each $Z_t$ separately to
$Z_t^*$ with law $P_n$, independent of its own $\mathcal G_t$,
with mismatch probability at most $C\varepsilon_n$. Such copies
can be constructed on a common extension; they need not be mutually
independent. The probability of any mismatch is at most
$Cn\varepsilon_n$. Conditional on its own fit, each squared
score difference at $Z_t^*$ has expectation at most $C\rho_n^2$.
The expectation of their average is therefore at most $C\rho_n^2$.
For $\rho_n>0$, Markov's inequality bounds the probability that
its square root exceeds $A\rho_n$ by $C/A^2$. Adding the mismatch
probability proves the assertion; if $\rho_n=0$, the coupled sum
is zero almost surely. Finally localization transfers all three
conclusions to the original fits.
\end{proof}

\section{Proofs of the inference results}\label{sec:supp-inference-proofs}

\begin{proof}[Proof of Theorem~\external{thm:ordinary} and Corollary~\external{cor:wald}]
Lemma~\ref{lem:replacement} gives the expansion
in \eqexternal{eq:linear}. For each fixed nonzero $z\in\mathbb R^d$,
the centered scalar array $z^{\T}\xi_{t,n}$ has uniformly bounded
$4+\delta$ moments and strong mixing coefficients bounded by
$C_\beta e^{-c_\beta h}$. The covariance envelope in
Lemma~\ref{lem:moments} gives
\[
 \var\left(n^{-1/2}\sum_tz^{\T}\xi_{t,n}\right)
 =z^{\T}\Omega_nz+O(n^{-1})\longrightarrow z^{\T}\Omega z>0.
\]
Geometric mixing ensures the weighted summability required by the
triangular-array central limit theorem of \citet{FrancqZakoian2005}.
Applying that theorem and the Cram\'er--Wold theorem, then using the
expansion, proves the Gaussian limit in \eqexternal{eq:linear}.

The oracle HAC estimate obeys
$\mathcal B_n(\phi(\nu_{0n}))=\mathcal B_n(\xi)\to\Omega$ by
Lemma~\ref{lem:moments}. Its trace is $\Op(1)$,
while \eqref{eq:rmsreplacement} bounds the feasible difference norm
by $\Op(\rho_n)$. Inequality~\eqref{eq:haccompare-sharp} therefore
bounds covariance replacement by
$\Op(\sqrt{L_n}\rho_n+L_n\rho_n^2)=\op(1)$.
Together with \eqref{eq:hacrate}, this gives the quantitative bound
\begin{equation}\label{eq:feasible-hac-rate}
 \norm{\widehat\Omega_n-\Omega_n}
 =\Op\left(\sqrt{L_n/n}+L_n^{-1}
             +\sqrt{L_n}\rho_n+L_n\rho_n^2\right).
\end{equation}
Thus $L_n/n\to0$ and $\sqrt{L_n}\rho_n\to0$ suffice for
consistency when $L_n\to\infty$. Positive definiteness of the limit
makes the inverse exist with probability tending to one and gives
$\widehat\Omega_n^{-1}\to\Omega^{-1}$.
Slutsky's theorem and the quadratic form of a normal vector give
the noncentral chi-squared law. Its distribution is continuous,
so the size and local rejection limits follow. For diverging
$\norm{\sqrt n\theta_n}$, the centered estimation term remains
tight and the eigenvalues of $\widehat\Omega_n$ are bounded above
in probability; the quadratic form therefore diverges in probability.
\end{proof}

\begin{proof}[Proof of Proposition~\external{prop:uniform}]
Choose any sequence $P_n\in\mathfrak P_n$. The uniform assumptions
make it an admissible triangular array, including simultaneous
localization and all required rate limits. Lemma~\ref{lem:moments}
bounds $\norm{\Omega_{n,P_n}}$ uniformly. Together with the lower
eigenvalue bound, this places these matrices in a compact subset of
the positive definite matrices. Every subsequence therefore has a
further subsequence on which $\Omega_{n,P_n}\to\Omega_*>0$.
Theorem~\external{thm:ordinary} and Corollary~\external{cor:wald}
give the Gaussian limit and feasible covariance consistency there.
Continuity of Cholesky factorization and its inverse, followed by
Slutsky's theorem, gives $T_{n,P_n}\Longrightarrow N_d(0,I_d)$.
The singular-covariance probability tends to zero as well. Since these
conclusions hold on a further subsequence of every subsequence, they
hold along the full selected sequence.

Weak convergence to the continuous Gaussian distribution implies
uniform convergence of the joint distribution functions: use a finite
grid on a large cube, coordinatewise monotonicity between grid points,
and marginal tail bounds outside the cube. If
the Gaussian approximation in \eqexternal{eq:uniform-gaussian} failed, a subsequence and choices
$P_n\in\mathfrak P_n$ would keep this distribution-function distance
bounded away from zero, contradicting the selected-sequence conclusion.
The same selection argument makes the singular-covariance probability
vanish uniformly. The continuous mapping theorem gives
$Q_{n,P_n}\Longrightarrow\chi_d^2$ for every selected sequence;
continuity of its distribution and the same argument prove
the Wald approximation in \eqexternal{eq:uniform-gaussian}. Finally, coverage is exactly the event
$Q_{n,P}\le\chi_{d,1-\alpha}^2$, including the stated convention
on the singular event. This proves \eqexternal{eq:uniform-coverage}.
\end{proof}

\begin{proof}[Proof of Proposition~\external{prop:raw-comparison}]
The local $L^2$ bound implies
$\norm{M_n(\eta)-\theta_n}\le Cr_n$ on $\mathcal E_n$.
Apply Lemma~\ref{lem:replacement} with
$\phi_t=g_t$, $\nu=\eta$, $\mathcal N_n=\mathcal E_n$ and
$\rho_n=B_n=r_n$. Its centered conclusion retains the sum of
population mean errors, yielding \eqexternal{eq:raw-expansion}
without requiring $\sqrt n r_n\to0$.
Its empirical squared-error bound and Lemma~\ref{lem:haccompare}
give covariance replacement error
$\Op(\sqrt{L_n}r_n+L_nr_n^2)=\op(1)$.
Lemma~\ref{lem:moments} gives oracle covariance consistency, so
$\widehat\Omega_n^g\to\Omega^g$ in probability. Positive definiteness
makes the inverse exist with probability tending to one and gives
$(\widehat\Omega_n^g)^{-1}\to(\Omega^g)^{-1}$.
The centered scores $\xi^g_{t,n}$ satisfy the same mixing and moment
conditions, with $\Omega_n^g\to\Omega^g>0$. The same application of
\citet{FrancqZakoian2005} and the Cram\'er--Wold theorem gives
$n^{-1/2}\sum_t\xi^g_{t,n}\Longrightarrow N_d(0,\Omega^g)$.
Together with \eqexternal{eq:raw-expansion}, Slutsky's theorem yields
$\sqrt n\widehat\theta_n^g\Longrightarrow N_d(h+\delta,\Omega^g)$.
The continuous mapping theorem proves \eqexternal{eq:raw-wald}.

Under the null, continuity of the noncentral chi-squared distribution
gives the rejection limit in \eqexternal{eq:raw-size}.
For strict overrejection when $\lambda_\delta>0$, rotate a standard
normal vector to write this law as
$(U_1+\sqrt{\lambda_\delta})^2+\sum_{i=2}^dU_i^2$, where the $U_i$
are independent standard normals. For $a>0$, the probability
$\pr\{|U_1+v|\le a\}$ strictly decreases as $v$ increases from zero:
its derivative is $\varphi(a+v)-\varphi(a-v)<0$ for $v>0$, where
$\varphi$ is the standard normal density. Conditioning on
$\sum_{i=2}^dU_i^2$ therefore gives
$F_{d,\lambda_\delta}(c)<F_{d,0}(c)$ for every $c>0$.
Taking $c=\chi_{d,1-\alpha}^2$ proves the strict inequality;
$\delta=0$ gives equality with $\alpha$.

For the random-drift qualification following the proposition, suppose
$(n^{-1/2}\sum_t\xi^g_{t,n},\Delta_n)\Longrightarrow(Z,D)$ and
$\sqrt n\theta_n\to h$. The same expansion and covariance consistency
give the limit
$(h+Z+D)^{\T}(\Omega^g)^{-1}(h+Z+D)$.
The marginal law of $Z$ is $N_d(0,\Omega^g)$, but no independence
of $Z$ and $D$ is implied.
\end{proof}

\subsection{Normalized comparisons between economic states}\label{sec:states}

Let $q_S,q_G$ be nonnegative, uniformly bounded state weights, and put
$\mu_{s,n}=\E q_{s,t}$ and $\Delta_{s,n}=\theta_{s,n}/\mu_{s,n}$.
For binary weights, $\Delta_{s,n}$ is the risk difference conditional
on $q_{s,t}=1$; general nonnegative weights give a normalized weighted
risk difference. Estimating $\mu_{s,n}$ changes the influence score. Define
\[
 \widehat\mu_s=n^{-1}\sum_tq_{s,t},\qquad
 \widehat\Delta_s=\widehat\theta_s/\widehat\mu_s,\qquad
 \widehat\kappa_{s,t}
   =\{\widehat\phi_{s,t}-\widehat\Delta_s q_{s,t}\}/\widehat\mu_s,
\]
with arbitrary fixed values when a sample denominator is zero. Write
$\boldsymbol\Delta_n=(\Delta_{S,n},\Delta_{G,n})^{\T}$ and
$\widehat{\boldsymbol\Delta}=(\widehat\Delta_S,\widehat\Delta_G)^{\T}$.
The
population influence score is
\[
 \kappa_{s,t,n}
   =\{\phi_{s,t,n}(\nu_{0n})-\Delta_{s,n}q_{s,t}\}/\mu_{s,n}.
\]

Suppose Assumptions~\external{ass:chronology}--\external{ass:learning}
and Assumption~\external{ass:array} hold for the two numerator scores,
with the rates in \eqexternal{eq:ordinaryrates} and the bandwidth
conditions of Corollary~\external{cor:wald}. Suppose also
$\inf_{n,s}\mu_{s,n}>0$ and $\sup_n\norm{\theta_n}<\infty$.
For a fixed full-row-rank matrix $C\in\mathbb R^{r\times2}$, let
$\Omega_{\kappa,n}$ be the long-run covariance of $\kappa_{t,n}$.
If $C\Omega_{\kappa,n}C^{\T}\to\Omega_C>0$, then
\[
 \sqrt n\{C\widehat{\boldsymbol\Delta}-C\boldsymbol\Delta_n\}
       \Longrightarrow N_r(0,\Omega_C),\qquad
 C\mathcal B_n(\widehat\kappa)C^{\T}
       \longrightarrow\Omega_C\quad\text{in probability}.
\]
Thus ordinary Wald inference applies to $C\boldsymbol\Delta_n$ using this covariance
estimate. Taking $C=(1,-1)$ compares the normalized risks across states.

\begin{proof}
Bounded weights and the covariance bound in Lemma~\ref{lem:moments}
give $\widehat\mu_s-\mu_{s,n}=\Op(n^{-1/2})$. Thus all sample
denominators are bounded away from zero with probability tending to
one, and $\boldsymbol\Delta_n$ is bounded. Write
$D_n=\operatorname{diag}(\mu_{s,n}^{-1})$ and
$\widehat D_n=\operatorname{diag}(\widehat\mu_s^{-1})$. Exactly,
\[
 \sqrt n(\widehat{\boldsymbol\Delta}-\boldsymbol\Delta_n)
  =\widehat D_nD_n^{-1}
    \left\{n^{-1/2}\sum_t\kappa_{t,n}
      +D_n n^{-1/2}\sum_t
          (\widehat\phi_t-\phi_{t,n}(\nu_{0n}))\right\}.
\]
Lemma~\ref{lem:replacement} makes the second sum $\op(1)$.
The first sum is $\Op(1)$, because $\kappa$ satisfies the same
mixing and moment bounds, and
$\widehat D_nD_n^{-1}=I+\Op(n^{-1/2})$. The centered scores
$C\kappa_{t,n}$ inherit these mixing and moment bounds, and their
long-run covariance converges to $\Omega_C>0$. Applying
\citet{FrancqZakoian2005} and the Cram\'er--Wold theorem to these
scores, as in the proof of Theorem~\external{thm:ordinary}, gives
the asserted Gaussian limit.

The same expansion gives
$\widehat{\boldsymbol\Delta}-\boldsymbol\Delta_n=\Op(n^{-1/2})$.
The sample denominators are bounded away from zero with probability
tending to one. Together with the empirical squared-error conclusion
of Lemma~\ref{lem:replacement} and sample second moments of order
$\Op(1)$, this implies
\[
 \left\{n^{-1}\sum_t
   \norm{\widehat\kappa_t-\kappa_{t,n}}^2\right\}^{1/2}
       =\Op(\rho_n+n^{-1/2}).
\]
Lemma~\ref{lem:moments} gives consistency of the oracle HAC estimate.
Applying \eqref{eq:haccompare-sharp} after multiplication by $C$
bounds its replacement error by
\[
 \Op\{\sqrt{L_n}(\rho_n+n^{-1/2})
             +L_n(\rho_n+n^{-1/2})^2\}=\op(1).
\]
This proves covariance consistency and hence the Wald conclusion.
\end{proof}

\section{Supporting results for forecasting implementations}
\label{sec:supp-learning}\label{sec:examples}\label{sec:learners}

We verify the correction and learning bounds for main-paper
Sections~\external{ex:lasso-implementation} and~\external{ex:neural-implementation}.
The population calculations specialize locally robust regression-function
inference \citep{ChernozhukovEtAl2022LR,ChernozhukovNeweySingh2022}.
The learning proofs use established concentration and empirical-risk
arguments, adapted to generated correction responses on the same training
window and simultaneous bounds over overlapping windows. They do not
provide optimization guarantees for a learning algorithm. The training
sets $J_k$ and averages $\E_k$ are defined in main-paper
Section~\external{sec:learner-summary}; tuning is deterministic unless
stated otherwise.

\subsection{Mean and projection corrections}
\label{sec:supp-lasso-proofs}

\begin{proposition}[Mean forecast correction]\label{prop:mean-correction}
Let $f_{j,0}=m_j:=\E(Y\mid X_j)$ and let the forecast space be
$L^2(\Ii_j,P)$.
Under Appendix Assumption~\external{ass:mean-technical}, the conditional-mean
representers and corrected score are \eqexternal{eq:mean-representer} and
\eqexternal{eq:mean-score}, respectively.
At the population arguments, the score identifies the target and is
jointly orthogonal. Let $L$ be the Lipschitz constant of the evaluation
derivative $\dot\ell(y,\cdot)$. For
$f_j,\alpha_{jr}\in L^2(\Ii_j,P)$, put
$u_j:=f_j-m_j$ and $b_{jr}:=\alpha_{jr}-\alpha_{jr,0}$. Then
\begin{equation}\label{eq:mean-bound}
 |\E\{\phi_r(f,\alpha)\}-\theta_{0r}|
 \le \frac{\norm{q_r}_\infty L}{2}\sum_j\norm{u_j}_2^2
                  +\sum_j\norm{b_{jr}}_2\norm{u_j}_2.
\end{equation}
If $q_r$ is measurable in $X_j$, the asymmetric loss gives
\begin{equation}\label{eq:asymmetric-mean-representer}
 \alpha_{jr,0}=2\epsilon_j\kappa q_r
                    \E\{(m_j-Y)_+\mid X_j\}.
\end{equation}
\end{proposition}

\begin{proof}[Proof of Proposition~\ref{prop:mean-correction}]
Conditional Jensen's inequality and the assumed square integrability
put $\alpha_{jr,0}$ in $L^2(\Ii_j,P)$. The Lipschitz evaluation
derivative and bounded instrument imply that
$q_r\dot\ell(Y,f_j)$ is square integrable for every $f_j\in L^2$.
The Taylor bound below and Cauchy--Schwarz then justify the
evaluation derivative representation and integrability of the loss
and correction products.
The half squared training score is $(f_j-Y)v$, with Hessian
$\E(uv)$. Conditional expectation gives
$\E\{\alpha_{jr,0}u_j\}
=\epsilon_j\E\{q_r\dot\ell(Y,m_j)u_j\}$,
which proves the representer equation. For each observation,
\[
 \ell(Y,m_j+u_j)-\ell(Y,m_j)
 =\dot\ell(Y,m_j)u_j+\zeta_j,\qquad
 |\zeta_j|\le L|u_j|^2/2.
\]
This inequality follows by integrating the Lipschitz derivative and
holds almost surely. In the correction difference, the terms involving
$b_{jr}(Y-m_j)$ have expectation zero. The linear loss terms cancel
$\alpha_{jr,0}u_j$ by the representer equation. The remaining mean is
$\sum_j\{\epsilon_j\E(q_r\zeta_j)-\E(b_{jr}u_j)\}$.
This proves \eqref{eq:mean-bound}, identification and joint
orthogonality. Conditional mean zero of $m_j-Y$ gives
\eqref{eq:asymmetric-mean-representer}.
\end{proof}

\begin{corollary}[Projection correction]\label{cor:projection}
Let $f_{j,0}=X_j^{\T}\beta_{j,0}$ be the least-squares projection of $Y$
on $X_j$, with positive-definite Gram matrix $\Sigma_j:=\E(X_jX_j^{\T})$,
as in Example~\external{ex:state-evaluation}.
Under Appendix Assumption~\external{ass:mean-technical}, the coefficient
representer satisfies \eqexternal{eq:projection-representer} componentwise,
with $q_r$ and $\alpha_{jr,0}$ in place of $q$ and $\alpha_{j,0}$.
The corrected score is \eqexternal{eq:mean-score}, with the prediction
and representer restricted to these linear forms. Its mean bound has the form
\eqref{eq:mean-bound}, with population prediction norms. Under squared
evaluation, putting
$u_j:=X_j^{\T}(\beta_j-\beta_{j,0})$ and
$b_{jr}:=X_j^{\T}(\alpha_{jr}-\alpha_{jr,0})$ gives the exact remainder
\[
 \E\{\phi_r(\beta,\alpha)\}-\theta_{0r}
       =\sum_j\{\epsilon_j\E(q_ru_j^2)-\E(b_{jr}u_j)\}.
\]
\end{corollary}

\begin{proof}[Proof of Corollary~\ref{cor:projection}]
The bounded instrument, Lipschitz derivative and stated integrability
give the Taylor bound in Proposition~\ref{prop:mean-correction}
at $f_{j,0}$, justifying differentiation of expected evaluation loss.
The Hessian of the training risk in coefficient directions is
$u^{\T}\Sigma_jv$. The representer equation gives the displayed
linear system in \eqexternal{eq:projection-representer}. The proof of
Proposition~\ref{prop:mean-correction}
then applies to coefficient directions: the expectation of a
representer perturbation times the residual is zero by the normal
equations. Squared evaluation has exact Taylor remainder $u_j^2$,
which gives the stated identity.
\end{proof}

\subsection{Simultaneous lasso bounds}

The argument allows any fixed finite number of rules and instrument
components. All constants are independent of $n$ and the window index.
For a scalar sequence $h=(h_t)$, define its empirical $L^2$ norm on
training window $J_k$ by
\[
 \|h\|_k:=\left\{\frac1{|J_k|}\sum_{t\in J_k}h_t^2\right\}^{1/2}.
\]
Write $\|U\|_{\psi_v}=\inf\{c>0:\E\exp(|U|^v/c^v)\le2\}$,
$v=1,2$. The distributional assumptions bound these norms as indicated.
For component $r$, the population and generated correction responses are
\[
 H_{jr,0}:=\epsilon_jq_r\dot\ell_\kappa(Y,X_j^{\T}\beta_{j,0}),
 \qquad
 \widehat H_{jrt,k}:=\epsilon_jq_{rt}\dot\ell_\kappa
            (Y_t,X_{jt}^{\T}\widehat\beta_{jk}).
\]
These are the componentwise versions of
\eqexternal{eq:lasso-fit}--\eqexternal{eq:lasso-representer}.

The next bound adapts the block-coupling argument of
\citet[Section~3]{DoukhanMassartRio1995} and Bernstein's inequality
\citep[Theorem~2.8.1]{Vershynin2018} to a growing collection of
overlapping training windows and sub-exponential functions.

\begin{lemma}[Concentration over overlapping windows]
\label{lem:window-concentration}
Let $\mathcal V_n$ contain at most $Cp_n^C$ centered measurable
functions with uniformly bounded sub-exponential norms. Under
geometric absolute regularity, with $r_n$ as in
\eqexternal{eq:lasso-rate} and $r_n\to0$,
\[
 \max_{k,V\in\mathcal V_n}|\E_k V|=O_{\Pr}(r_n),
\]
and a sufficiently large fixed multiple of $r_n$ bounds the maximum
with probability tending to one. This includes triangular arrays of
bounded indicator functions.
\end{lemma}

\begin{proof}[Proof of Lemma~\ref{lem:window-concentration}]
In each window form consecutive blocks of length
$v_n=\lceil C_v\log(enK_nT_n)\rceil$; the final block may be shorter.
Within each parity, successive blocks have separation at least $v_n$.
Berbee coupling replaces them by independent blocks with total
failure probability at most $CK_nT_n\exp(-c_\beta v_n)$.
Couple whole observations, so the event serves every function in
$\mathcal V_n$. For a copied block sum $W_{i,V}$,
$\|W_{i,V}\|_{\psi_1}\le Cv_n$, and the sum of their squared
norms within a parity is at most $CT_nv_n$. The independent
sub-exponential Bernstein inequality
\citep[Theorem 2.8.1]{Vershynin2018} and a union bound yield,
for $0<t\le1$,
\[
 \Pr\{\max_{k,V}|\E_kV|>t\}
 \le CK_np_n^C\exp(-cT_nt^2/v_n)
           +CK_nT_n\exp(-c_\beta v_n).
\]
Take $C_v$ sufficiently large and $t=C_0r_n$ with $C_0$
sufficiently large. Both terms tend to zero. Different windows
need not be independent.
\end{proof}

The following deterministic argument extends the lasso bounds in
\citet[Theorem~1 and Appendix~A]{WongLiTewari2020} to an additive
response error controlled in empirical $L^2$. This permits correction
responses generated on the forecast training sample.

\begin{lemma}[Lasso with generated responses]\label{lem:generated-lasso}
Consider a sequence of deterministic regression problems, with the
sequence index suppressed. For a sample of size $T$, let
$X\in\R^{T\times p}$ and consider the response $Xa_0+\zeta+d$,
where $a_0\in\R^p$ has at most $s$ nonzero entries and
$\zeta,d\in\R^T$.
Write $\|v\|_T^2:=T^{-1}\sum_{t=1}^T v_t^2$ and let
$\widehat a$ minimize
$\tfrac12\|Xa_0+\zeta+d-Xa\|_T^2+\lambda\norm a_1$, with
$\lambda>0$. Suppose
$\|X^{\T}\zeta/T\|_\infty\le\lambda/2$.
Put $\epsilon=\|d\|_T$,
$z=\|\widehat a-a_0\|$ and
$t=\|X(\widehat a-a_0)\|_T$.
If $\lambda_{\min}(\Sigma)\ge c>0$ and
$\|X^{\T}X/T-\Sigma\|_{\max}\le\tau$, with
$\tau s=o(1)$ and $\tau\epsilon^2/\lambda^2=o(1)$, then eventually
\[
 z+t\le C(\sqrt s\,\lambda+\epsilon),\qquad
 \|\widehat a-a_0\|_1
    \le C(s\lambda+\sqrt s\,\epsilon+\epsilon^2/\lambda).
\]
The perturbation $d$ may depend on the entire sample.
\end{lemma}

\begin{proof}[Proof of Lemma~\ref{lem:generated-lasso}]
Let $S$ be the support of $a_0$ and $\Delta=\widehat a-a_0$.
The basic inequality, the noise bound and Cauchy--Schwarz give
\[
 t^2+\lambda\|\Delta_{S^c}\|_1
                  \le3\lambda\|\Delta_S\|_1+2\epsilon t.
\]
Hence
$t^2\le6\lambda\sqrt s\,z+4\epsilon^2$ and
$\|\Delta\|_1\le4\sqrt s\,z+2\epsilon^2/\lambda$.
The Gram bound implies
\[
 cz^2\le t^2+\tau\|\Delta\|_1^2
 \le6\lambda\sqrt s\,z+4\epsilon^2
                         +32\tau sz^2+8\tau\epsilon^4/\lambda^2.
\]
Since $\tau s=o(1)$, absorb $32\tau sz^2$ into the left-hand side.
Also, $\tau\epsilon^2/\lambda^2=o(1)$ bounds
$8\tau\epsilon^4/\lambda^2$ by $o(1)\epsilon^2$. Using
$6\lambda\sqrt s\,z\le cz^2/4+C s\lambda^2$ then gives the bound on
$z$. Substitution into the preceding inequalities gives the bounds on
$t$ and $\|\widehat a-a_0\|_1$.
\end{proof}

\begin{proof}[Proof of Corollary~\external{cor:lasso}]
The projection property bounds $\|X_j^{\T}\beta_{j,0}\|_2$
by $\|Y\|_2$. The evaluation derivative has linear growth in
$|Y|+|X_j^{\T}\beta_{j,0}|$.
The representer is the linear projection of its signed,
instrumented derivative, so its prediction also has bounded $L^2$
norm. The lower Gram bound makes both coefficient norms bounded.
The directional sub-Gaussian assumption therefore bounds all their
prediction moments. Products of predictor coordinates with the
population forecast residual $Y-X_j^{\T}\beta_{j,0}$ or the population
correction residual $H_{jr,0}-X_j^{\T}\alpha_{jr,0}$ are uniformly
sub-exponential.

Apply Lemma~\ref{lem:window-concentration} to centered Gram entries,
the forecast score $X_j(Y-X_j^{\T}\beta_{j,0})$, and the correction
score $X_j\{H_{jr,0}-X_j^{\T}\alpha_{jr,0}\}$.
The score means are zero by the two population equations.
With probability tending to one their coordinate maxima are bounded
by $\lambda_n/2$ and the Gram error is at most $Cr_n$.
Lemma~\ref{lem:generated-lasso} with $\epsilon=0$ proves the
forecast coefficient and empirical prediction bounds $Ca_n$.

The derivative in \eqexternal{eq:asymmetric-loss} is Lipschitz with constant
$2(1+\kappa)$. Thus the empirical norm of
$\widehat H_{jr,k}-H_{jr,0}$ is at most $Ca_n$, on the same
event. Apply the generated response lemma with
$\tau=Cr_n$, $\lambda_n\asymp r_n$, $s=s_{a,n}$ and
$\epsilon\le Ca_n$. Its two smallness conditions follow from
$(1+s_{f,n}+s_{a,n})r_n\to0$, which also ensures $r_n\to0$
for the concentration lemma when both sparsities vanish. The resulting
correction coefficient and empirical prediction errors are bounded
by $Cc_n$ simultaneously over rules, components and windows.
For any fixed $p$, the directional sub-Gaussian bound implies
$\|X_j^{\T}v\|_p\le C_p\|v\|$ for each fixed coefficient vector;
holding the realized vector fixed gives the population prediction
claims.

The Taylor argument in Corollary~\ref{cor:projection} bounds the
mean error by $C(a_n^2+a_nc_n)$.
The loss difference is bounded by a constant times
$(|Y|+|f_{j,0}|+|u_j|)|u_j|$. The correction difference is a sum
of products of prediction and representer errors with residuals
and true representers. H\"older's inequality and the directional
moment bounds therefore give score $L^2$ error $C(a_n+c_n)$
on deterministic coefficient neighborhoods containing all fits.
The true score is a sum of bounded multiples of products of
sub-Gaussian variables and has moments of every fixed order.
These facts verify the hypotheses of the inference results.
\end{proof}

Expanding the squared loss in \eqexternal{eq:lasso-representer}
shows that the correction fit is also a penalized empirical Riesz
regression in the sense of
\citet[Section~3.2]{ChernozhukovEtAl2024RieszRegression}.
The next result permits approximation error in both coefficient vectors
while retaining the population projection and correction targets.

\begin{proposition}[Approximately sparse forecasts and corrections]
\label{prop:lasso-approximate}
Maintain Appendix Assumption~\external{ass:lasso-technical}, except
replace its exact-sparsity requirements and final sparsity condition by
the following conditions. For fixed $q_f,q_a\in(0,1)$ and deterministic
$R_{f,n},R_{a,n}\ge0$, suppose, for every rule $j$ and component $r$,
\[
 \sum_{\ell=1}^{p_n}|\beta_{j,0,\ell}|^{q_f}\le R_{f,n},
 \qquad
 \sum_{\ell=1}^{p_n}|\alpha_{jr,0,\ell}|^{q_a}\le R_{a,n}.
\]
Assume also
\[
 r_n\to0,\quad
 R_{f,n}r_n^{1-q_f}+R_{a,n}r_n^{1-q_a}\to0.
\]
Set
\[
 a_n:=\sqrt{R_{f,n}}\,r_n^{1-q_f/2},\qquad
 c_n:=a_n+\sqrt{R_{a,n}}\,r_n^{1-q_a/2}.
\]
With probability tending to one, simultaneously over rules,
components and training windows,
\begin{align*}
 \|\widehat\beta_{jk}-\beta_{j,0}\|
  +\|X_j^{\T}(\widehat\beta_{jk}-\beta_{j,0})\|_k
  &\le Ca_n,\\
 \|\widehat\alpha_{jr,k}-\alpha_{jr,0}\|
  +\|X_j^{\T}(\widehat\alpha_{jr,k}-\alpha_{jr,0})\|_k
  &\le Cc_n.
\end{align*}
The localization bounds hold with $B_n=C(a_n^2+a_nc_n)$ and
$\rho_n=C(a_n+c_n)$. Theorem~\external{thm:ordinary} and
Corollary~\external{cor:wald} therefore apply under their remaining
covariance, chronology, bandwidth and rate conditions.
\end{proposition}

\begin{proof}[Proof of Proposition~\ref{prop:lasso-approximate}]
The projection equations and lower Gram bound give uniformly bounded
population coefficient norms, as in the proof of
Corollary~\external{cor:lasso}; this argument does not use exact
sparsity. Lemma~\ref{lem:window-concentration} therefore gives the same
simultaneous event on which both population residual scores are bounded
by $\lambda_n/2$ in maximum norm and
\[
 \max_{j,k}\|\widehat\Sigma_{jk}-\Sigma_j\|_{\max}\le Cr_n,
 \qquad \widehat\Sigma_{jk}:=\E_k(X_jX_j^{\T}).
\]
The directional sub-Gaussian condition also gives
$\sup_{j,n}\|\Sigma_j\|_{\mathrm{op}}<\infty$.

For any deterministic coefficient vector $b$ satisfying
$\sum_\ell|b_\ell|^q\le R$, retain its coordinates with
$|b_\ell|>r_n$ to obtain $b^*$. Its support size and tail satisfy
\[
 |\operatorname{supp}(b^*)|\le Rr_n^{-q},\qquad
 \|b-b^*\|^2\le Rr_n^{2-q},\qquad
 \|b-b^*\|_1\le Rr_n^{1-q}.
\]
On the Gram event, writing $v=b-b^*$ gives, uniformly over windows,
\[
 \|X_j^{\T}v\|_k^2
 \le C\|v\|^2+Cr_n\|v\|_1^2
 \le CRr_n^{2-q}\{1+Rr_n^{1-q}\}.
\]
For either coefficient family, the expression in braces is eventually
bounded by the assumed smallness condition.

Apply this construction to $\beta_{j,0}$ and write the training
response as
\[
 Y=X_j^{\T}\beta_{j,0}^*
       +(Y-X_j^{\T}\beta_{j,0})
       +X_j^{\T}(\beta_{j,0}-\beta_{j,0}^*).
\]
The middle term is the true projection residual, whose empirical score
is bounded on the concentration event. The last term has empirical
norm at most $Ca_n$. In Lemma~\ref{lem:generated-lasso}, take
$s\le R_{f,n}r_n^{-q_f}$, $\tau=Cr_n$ and
$\lambda_n\asymp r_n$. Both $\tau s$ and
$\tau\epsilon^2/\lambda_n^2$ are bounded by a constant times
$R_{f,n}r_n^{1-q_f}\to0$. The lemma and the coefficient and empirical
tail bounds give the asserted forecast rates relative to
$\beta_{j,0}$.

For the correction, use the sparse approximation
$\alpha_{jr,0}^*$ and decompose its response as
\begin{align*}
 \widehat H_{jr,k}
 &=X_j^{\T}\alpha_{jr,0}^*
   +(H_{jr,0}-X_j^{\T}\alpha_{jr,0})\\
 &\quad+X_j^{\T}(\alpha_{jr,0}-\alpha_{jr,0}^*)
   +(\widehat H_{jr,k}-H_{jr,0}).
\end{align*}
The second term is the population correction residual; its empirical
score is bounded by $\lambda_n/2$ in maximum norm on the concentration
event. The sum of the last two terms has empirical $L^2$ norm
$\|\cdot\|_k$ at most $Cc_n$. The approximation tail is controlled by the
preceding argument, and the generated response error by the Lipschitz
evaluation derivative and the forecast prediction bound.
The two smallness requirements in Lemma~\ref{lem:generated-lasso}
now follow from
\[
 \tau s\le CR_{a,n}r_n^{1-q_a}\to0,\qquad
 \frac{\tau\epsilon^2}{\lambda_n^2}
 \le C\{R_{f,n}r_n^{1-q_f}+R_{a,n}r_n^{1-q_a}\}\to0.
\]
The lemma and the tail bounds give the correction rates. Finally,
$a_n,c_n\to0$, and the coefficient-neighborhood moment argument in the
proof of Corollary~\external{cor:lasso} gives the stated mean and
$L^2$ score bounds without using exact sparsity.
\end{proof}

\subsection{Neural probability forecasts}
\label{sec:nn-example}\label{sec:nn-proofs}

For the cost criterion and analytic score in
\eqexternal{eq:nn-representer}--\eqexternal{eq:nn-score}, the following
bounds transfer prediction accuracy directly to score accuracy.

\begin{proposition}[Cost-weighted probability-risk correction]\label{prop:nn-correction}
The score in \eqexternal{eq:nn-score} identifies
$\theta_0=\E\{\ell_\tau(Y,p_{1,0})-\ell_\tau(Y,p_{2,0})\}$.
Its joint prediction--representer form is Neyman orthogonal at the
population targets. For fixed measurable
$p_j\in[0,1]$, put $a_j:=\norm{p_j-p_{j,0}}_2$. Then
\begin{equation}\label{eq:nn-mean-bound}
 |\E\phi(p)-\theta_0|\le(1+2\tau)\sum_j a_j^2,\qquad
 \norm{\phi(p)-\phi(p_0)}_2\le2(1+2\tau)\sum_j a_j.
\end{equation}
Moreover $\norm{A_\tau(p_j)-A_\tau(p_{j,0})}_2\le2\tau a_j$.
At $\tau=0$, the correction is identically zero.
\end{proposition}

\begin{proof}[Proof of Proposition~\ref{prop:nn-correction}]
Half squared training loss on $L^2(\Ii_j,P)$ has directional score
$(p_j(X_j)-Y)v(X_j)$ and Hessian $\E(uv)$. Conditional expectation gives
\[
 \E\{\partial_p\ell_\tau(Y,p_{j,0})\mid X_j\}
 =-2\tau p_{j,0}(1-p_{j,0})=A_\tau(p_{j,0}).
\]
Together with $\E(Y-p_{j,0}\mid X_j)=0$, this proves the signed
representer equation and identification. For the one-rule score
$\ell_\tau(Y,p_j)+a(X_j)(Y-p_j)$, with a separate correction function
$a$, the two conditional identities cancel the prediction and correction
derivatives at $(p_{j,0},A_\tau(p_{j,0}))$, respectively.
Thus orthogonality also holds when prediction and representer are
estimated separately.

For the quantitative bounds, define
$h_\tau(p):=p(1-p)\{1+\tau(1-p)\}$.
Since $Y^2=Y$, direct expansion gives the exact identity
\begin{equation}\label{eq:nn-one-step}
 H_\tau(Y,p)=h_\tau(p)+h_\tau'(p)(Y-p).
\end{equation}
For fixed $p_j$, conditional expectation replaces $Y$ by $p_{j,0}$.
Writing $e_j=p_j-p_{j,0}$ and expanding the cubic polynomial
$h_\tau(p)=(1+\tau)p-(1+2\tau)p^2+\tau p^3$ yields
\[
 \E\phi(p)-\theta_0
 =\sum_j\epsilon_j\E\left[
 (1+2\tau-3\tau p_{j,0})e_j^2-2\tau e_j^3\right].
\]
Alternatively, Taylor's theorem about $p_j$ and
$\sup_{[0,1]}|h_\tau''|\le2(1+2\tau)$ bound each conditional remainder
by $(1+2\tau)e_j^2$. Taking expectations proves the first inequality in
\eqref{eq:nn-mean-bound}. Differentiating
\eqref{eq:nn-one-step} gives
$\partial_pH_\tau(y,p)=h_\tau''(p)(y-p)$, whose absolute value is at most
$2(1+2\tau)$ for binary $y$ and $p\in[0,1]$.
The mean-value inequality and Minkowski's inequality prove the second
inequality in \eqref{eq:nn-mean-bound}. Finally,
$|A_\tau'(p)|\le2\tau$ gives the representer-error bound, and
$A_0=0$ proves the aligned-loss assertion.
\end{proof}

\begin{proof}[Proof of Corollary~\external{cor:nn-rates}]
The prediction bound follows from a relative concentration inequality
for squared-loss excess risks. Suppress the rule and row indices, and
write $P$ for population expectation and $P_k=\E_k$ for a training
window's empirical average. For deterministic $p\in\mathcal P_{j,n}$,
put
\[
 D_p=(Y-p(X_j))^2-(Y-p_{j,0}(X_j))^2,
 \qquad \mu_p=PD_p=\norm{p-p_{j,0}}_2^2.
\]
The equality for $\mu_p$ follows by conditioning on $X_j$.
Bounded outcomes and predictions give $|D_p|\le2|p-p_{j,0}|$ and
$PD_p^2\le4\mu_p$.

Split each training window into consecutive blocks of length $v_n$,
and group odd and even blocks separately. Since $r_{N,n}\to0$,
$2v_n\le T_n$ eventually. For each parity, absolute regularity permits
coupling the blocks to independent copies, with failure probability
at most $C(T_n/v_n)\beta_n(v_n)$ per window. The coupling concerns the
data blocks themselves, so its failure probability is not multiplied
by the number of functions in a cover. Across all windows it is at most
$CK_nT_n\beta_n(v_n)$, which tends to zero for sufficiently large $C_v$.

A block average of $D_p$ is bounded by 2. Jensen's inequality bounds
its variance by $4\mu_p$, without requiring independence within the
block. Bernstein's inequality \citep[Theorem~2.8.4]{Vershynin2018}
for the coupled independent block averages therefore bounds each parity's empirical deviation by
\[
 C\sqrt{\mu_p v_n u/T_n}+Cv_nu/T_n
\]
outside an event of probability $2e^{-u}$. Incomplete end blocks have
at most $2v_n$ observations and contribute at most $Cv_n/T_n$.
Choose class representatives from a $T_n^{-2}$-cover for each rule,
giving a $2T_n^{-2}$-net of the same cardinality, and use a union bound over
both parities, both rules, all windows and all net elements, with
$u=H_n+C\log(enK_n)$. Put
$V_n=v_n\{H_n+\log(enK_n)\}/T_n$.
Young's inequality then yields
\begin{equation}\label{eq:nn-relative}
 |(P_k-P)D_p|\le\tfrac14\mu_p+C(V_n+T_n^{-2})
 \quad\text{for every }p\in\mathcal P_{j,n},j,k
\end{equation}
with probability tending to one. To extend from the net, use
$|D_p-D_{\widetilde p}|\le2\norm{p-\widetilde p}_\infty$ and
$|\mu_p-\mu_{\widetilde p}|\le2\norm{p-\widetilde p}_\infty$.
The bound is simultaneous over windows; their overlap poses no
additional independence requirement.

On this event and the simultaneous optimization event, compare
$\widehat p_{jk}$ with a deterministic $p_*$ whose population error
approaches the class infimum. Applying \eqref{eq:nn-relative} to both
predictors and using \eqexternal{eq:nn-optimization} gives
\[
 \tfrac34\norm{\widehat p_{jk}-p_{j,0}}_2^2
 \le\tfrac54\norm{p_*-p_{j,0}}_2^2
       +\varepsilon_n+2C(V_n+T_n^{-2}).
\]
Take the infimum and then square roots, simultaneously over $j,k$,
to obtain \eqexternal{eq:nn-rate}. The mean and score bounds in
Proposition~\ref{prop:nn-correction} give $B_n=Cr_{N,n}^2$ and $\rho_n=Cr_{N,n}$ on the resulting
deterministic prediction neighborhoods. Binary outcomes and bounded
predictions make the oracle score uniformly bounded. Apply
Theorem~\external{thm:ordinary} and Corollary~\external{cor:wald} to
obtain feasible inference under the stated remaining conditions.
\end{proof}

\paragraph{Neural covering numbers.}
For each $n$, consider an architecture with $N_{\mathrm{par},n}$
parameters in $[-M_n,M_n]$, compact predictor support and Lipschitz
activations. Its prediction map is Lipschitz in the parameter maximum
norm, uniformly over the predictor support, with a finite constant $C_n$.
To see this, bound the input to the first affine layer on the compact
support. Each affine layer has bounded coefficients and each activation
is Lipschitz; induction therefore bounds both its output and its change
under a perturbation of the parameter vector in maximum norm.
Applying a Lipschitz output map with range in $[0,1]$ preserves the
property. A parameter grid of spacing $T_n^{-2}/C_n$ consequently gives
\[
 H_n\le N_{\mathrm{par},n}\log\{2+2M_nC_nT_n^2\}.
\]
The constant $C_n$ must include width and depth dependence. This
argument permits fitted hidden-layer weights and applies to both smooth
and ReLU activations.

\begin{proposition}[Training-only selection of neural weight decay]
\label{prop:nn-validation}
Let $p_\vartheta$ have $N_{\mathrm{par},n}$ parameters in $[-3,3]$, with $i_0$ indexing
its affine output intercept. For each rule, let the declared class be
the image of this bounded parameter set, and define
$R_n(\vartheta)=\tfrac12\sum_{i\ne i_0}\vartheta_i^2$.
A training-data selection rule chooses
$0\le\widehat\lambda\le\kappa_{\max}/T_n$, for fixed
$\kappa_{\max}<\infty$, before refitting on all $T_n$ observations.
Suppose the final refit belongs to the declared class and, simultaneously
over rules and training windows,
\[
 \tfrac12\E_k(Y-p_{\widehat\vartheta})^2
   +\widehat\lambda R_n(\widehat\vartheta)
 \le\inf_\vartheta\left\{
     \tfrac12\E_k(Y-p_\vartheta)^2
          +\widehat\lambda R_n(\vartheta)\right\}
          +\varepsilon_{\mathrm{pen},n}
\]
with probability tending to one. Then \eqexternal{eq:nn-optimization}
holds with
\begin{equation}\label{eq:nn-penalty-gap}
 \varepsilon_n=2\varepsilon_{\mathrm{pen},n}
       +9\kappa_{\max}(N_{\mathrm{par},n}-1)/T_n.
\end{equation}
\end{proposition}

\begin{proof}
Fix the training observations and their selected penalty.
The parameter bound gives
$0\le R_n(\vartheta)\le9(N_{\mathrm{par},n}-1)/2$.
Compare the final penalized objective with a parameter approaching
the infimum of the unpenalized empirical squared loss. Dropping the
nonnegative penalty at the final fit gives
\[
 \tfrac12\E_k(Y-p_{\widehat\vartheta})^2
 \le\tfrac12\inf_\vartheta\E_k(Y-p_\vartheta)^2
      +\tfrac92\widehat\lambda(N_{\mathrm{par},n}-1)
      +\varepsilon_{\mathrm{pen},n}.
\]
Multiply by two and use
$\widehat\lambda\le\kappa_{\max}/T_n$ to obtain
\eqref{eq:nn-penalty-gap}. The argument is samplewise for every
permitted selected penalty; independence between selection and refitting
is unnecessary.
\end{proof}

For penalties satisfying Proposition~\ref{prop:nn-validation}, the
additional prediction-error term is $O(\sqrt{N_{\mathrm{par},n}/T_n})$.
Together with the approximation and entropy requirements, sufficient
conditions for the mean remainder include
$\sqrt n\,\varepsilon_{\mathrm{pen},n}\to0$ and
$\sqrt n\,N_{\mathrm{par},n}/T_n\to0$; for $T_n\asymp n$ the latter is
$N_{\mathrm{par},n}=o(\sqrt n)$. The covariance and chronological learning
restrictions apply to the complete rate $r_{N,n}$. These conditions
preserve the conditional probability target while allowing a vanishing
penalty. The bound on the gap from the global minimum remains an
assumption; validation based on prediction accuracy alone does not
establish the learning rates needed for valid inference.

\section{Covariance degeneracy and ordinary studentization}
\label{sec:degeneracy}

For a centered stationary score $\xi_t$, absolute summability of
$\Gamma(h):=\cov(\xi_0,\xi_h)$ gives
\[
 \Omega:=\sum_{h\in\mathbb Z}\Gamma(h)
   =\lim_{N\to\infty}N^{-1}\var\left(\sum_{t=1}^N\xi_t\right)
   \succeq0.
\]
Indeed, the finite-sample expression is
$\Gamma(0)+\sum_{h=1}^{N-1}(1-h/N)\{\Gamma(h)+\Gamma(h)^{\T}\}$;
dominated convergence gives the equality and preserves positive
semidefiniteness. For iid scores or martingale differences,
$\Omega=\var(\xi_t)$. Under dependence, positive marginal variance
need not imply positive long-run variance: if $E_t$ is iid with
variance one and $\xi_t=E_t-E_{t-1}$, then
$\var(\xi_t)=2$ but
$N^{-1}\var(\sum_{t=1}^N\xi_t)=2/N\to0$.

The positive corrected variance in main-paper
Example~\external{ex:asymmetric-evaluation} also occurs under
geometric absolute regularity. Let $P$ be that example's observation
law and use the stationary transition kernel
$K(z,\cdot)=\varrho\delta_z+(1-\varrho)P$, where $0<\varrho<1$.
A refresh couples the future to an independent stationary version,
so $\beta(h)\le\varrho^h$. For a centered score $v(Z)$, iterating
the kernel gives
$\E\{v(Z_h)\mid Z_0\}=\varrho^hv(Z_0)$.
Its long-run covariance is therefore
$(1+\varrho)/(1-\varrho)$ times its marginal covariance.

The next example shows why the main theorem requires a nondegenerate
corrected-score covariance even when the quadratic mean-error rate holds.

\begin{example}[Failure of ordinary studentization]\label{ex:failure}
Let $X$ take values $-1$ and $1$ with equal probability, and let
$Y\sim N(0,\sigma_\varepsilon^2)$, with $\sigma_\varepsilon>0$,
be independent of $X$.
Compare the zero forecast with $x\mapsto\beta x$, trained by least
squares and evaluated with squared loss. Both population forecasts
and their representers are zero; use the zero correction.
From $T$ iid training observations, estimate
$\widehat\beta=T^{-1}\sum_{s=1}^T X_sY_s$ and evaluate it on an
independent sample of $n$ iid observations, with
$n/T\to\omega\in(0,\infty)$.
The fitted loss contrast and its empirical variance are
\[
 g_t=2\widehat\beta X_tY_t-\widehat\beta^2,\qquad
 \widehat\sigma_g^2=n^{-1}\sum_t(g_t-\bar g)^2.
\]
For independent standard normal $A,B$,
\[
 T\bar g\ \Longrightarrow\
 \sigma_\varepsilon^2\{2AB/\sqrt\omega-A^2\},\qquad
 \frac{\sqrt n\bar g}{\widehat\sigma_g}
 \ \Longrightarrow\
 \operatorname{sign}(A)B-\frac{\sqrt\omega}{2}|A|.
\]
The second limit is not standard normal, and its square is not
$\chi_1^2$. Replacing $\widehat\sigma_g^2$ by the Bartlett estimate
$\mathcal B_n(g)$ gives the same limit when $L_n\to\infty$ and
$L_n/n\to0$.
\end{example}

\begin{proof}
The variables $\sqrt T\widehat\beta/\sigma_\varepsilon$ and
$n^{-1/2}\sum_tX_tY_t/\sigma_\varepsilon$ are independent standard
normals. Also,
\[
 \widehat\sigma_g^2=4\widehat\beta^2
       n^{-1}\sum_t(X_tY_t-\overline{XY})^2,
 \qquad
 n^{-1}\sum_t(X_tY_t-\overline{XY})^2\to\sigma_\varepsilon^2
\]
in probability. Substitution and the continuous mapping theorem,
using $\pr(A=0)=0$, give both limits. The second limiting variable
has mean $-\sqrt\omega\,\E|A|/2<0$ and second moment
$1+\omega/4$, proving both distributional assertions.
For Bartlett studentization, affine equivariance gives
$\mathcal B_n(g)=4\widehat\beta^2\mathcal B_n(XY)$, and
Lemma~\ref{lem:moments} gives
$\mathcal B_n(XY)\to\sigma_\varepsilon^2$ in probability.
The same substitution proves the final assertion.
\end{proof}

Here $\widehat\beta=\Op(n^{-1/2})$, so the localization radius
$a_n=\log(en)/\sqrt n$ contains its error with probability tending
to one and satisfies $\sqrt n a_n^2\to0$.
The oracle score is identically zero, and its covariance therefore
fails the nondegeneracy condition of ordinary Wald inference.

\clearpage
\section{Simulation designs and supporting evidence}\label{app:simulations}
\setcounter{table}{0}
\renewcommand{\thetable}{\thesection.\arabic{table}}

\subsection{Design, chronology and comparisons}\label{app:sim-designs}

This section specifies the designs underlying the main results and examines
size robustness, local power and the sources of remaining distortion.
Both experiments compare population evaluation risks. The lasso design
varies the number of regressors, training sample size and refitting frequency.
The neural design varies training information for a fixed architecture
applied to nonlinear event probabilities, with an analytic correction.

The lasso study comprises 111 configurations and 77,250 replications:
1,000 per standard null configuration, 500 per local alternative, and
250 per additional null configuration with refitting at every date. The neural study has three null
configurations with 1,024 replications each and eight alternatives with 512
each, giving 7,168 replications. Pilots and superseded experiments are
excluded. At a true rejection probability of 5\%, Monte Carlo standard
errors are 0.69 percentage points for standard lasso nulls, 1.38 for
refitting at every date and 0.68 for neural nulls.

\paragraph{Chronology.}
Both designs are stationary, with dependence parameter $\varrho=0.5$
and the transition laws specified below. The evaluation sample has
$n\in\{500,1000\}$ observations; each refit uses $T$ observations
available before its evaluation block.
Lasso comparisons use $T/n\in\{0.5,1,2\}$, with longer histories only in
an analytic-penalty diagnostic. Neural nulls use
$(n,T)=(500,4000),(1000,4000),(1000,8000)$; neural power uses
$(n,T)=(1000,4000)$. Lasso forecasts are estimated once, five times, every 25 evaluation
dates, or at every date in the additional $n=T=500$ null experiment.
Neural forecasts are estimated five times. Training windows overlap,
and an evaluation outcome can enter a subsequent training window after
the required temporal separation. The gaps are
\[
 b_n^{\rm L}=\left\lceil\frac{2\log(en)}{|\log\varrho|}\right\rceil,
 \qquad
 b_n^{\rm N}=\left\lceil\frac{2\log n}{|\log\varrho|}\right\rceil.
\]
The primary Bartlett bandwidth is
$L_n=\max\{\lceil n^{1/4}\rceil,b_n\}$, using the respective gap.
All tuning uses eligible training observations only.

Comparable configurations for a given learner and sample size share latent
paths and evaluation dates. Shorter histories use suffixes of the same past;
larger lasso specifications add irrelevant regressors. Alternatives of
opposite signs share latent randomness, although their outcomes may differ.
The two long-history lasso penalty checks use a separate common path.
Pairing improves the precision of method comparisons while preserving
each forecast's information set.

\paragraph{Wald and bootstrap tests.}
We compare Raw, Bootstrap Raw, Orthogonal and Bootstrap Orthogonal.
The Wald tests use estimated forecasts and corrections, the respective
Bartlett covariance estimates, and the two-sided normal critical value
$z_{0.975}$. Coverage refers to the population risk contrast; rejection
frequencies concern the null that this contrast is zero.

Bootstrap Raw and Bootstrap Orthogonal center their respective evaluation
scores and resample circular blocks of length $2(L_n+1)$, with 499 resamples
for lasso and 999 for neural forecasts. Each resample recomputes the
studentized mean and Bartlett covariance. For $B$ resamples, the reported
$p$-value is
\[
 \widehat p=\frac{1+\sum_{b=1}^{B}\ind\{|t_b^*|\ge |t|\}}{B+1}.
\]
Raw and corrected scores use the same resampling indices. Holding fitted forecast and correction functions fixed across bootstrap
resamples isolates the effect of replacing normal critical values by
bootstrap critical values. The bootstrap does not reestimate
the learners, and its validity is not implied by the Gaussian approximation
established in the paper.

Studentization requires a finite variance estimate exceeding $10^{-24}$;
the bootstrap applies this requirement to every resample. All displayed
CV-lasso and neural statistics satisfy it, and no simulation run fails.
Degenerate statistics are reported separately from nonrejections, including
those arising under the analytic penalty in Section~\ref{app:sim-support}.
Learning and optimization diagnostics are not used to exclude replications.

\subsection{Lasso design and forecast--evaluation mismatch}\label{app:sim-lasso}

At stationarity, $S_1,S_2,W$ and the irrelevant regressors are
independent symmetric signs. Their joint vector is retained with probability
$\varrho$ and otherwise refreshed independently, giving
$\beta(h)\le\varrho^h$. Independent outcome noise is uniform on $[-1,1]$.
The outcome and evaluation state are
\[
 Y_t=(b_1+cW_t)S_{1t}+b_2S_{2t}+\varepsilon_t,
 \qquad b_1=0.3,\quad c=0.6,\quad q_t=\ind\{W_t=1\}.
\]
Each regressor vector contains an intercept, its own signal $S_j$, $W$, and
irrelevant signs; it excludes the other signal and the interaction terms.
The baseline dimension is $p=32$ at both sample sizes. Varying
$p\in\{32,128,256\}$ changes complexity while holding sparsity and the
population targets fixed. Across the comparison, $p/T$ ranges from
0.016 to 0.512, with no configuration having $p>T$.
The population Gram matrix is the identity, and the training projections
are $f_{j,0}=b_jS_j$. Evaluation uses $q(Y-f)^2$.
The omitted interaction gives the first rule a nonzero expected evaluation
derivative, as in main-paper Example~\external{ex:state-evaluation}.

Write $a_j$ for the rule's representer before applying the contrast sign,
so that the main paper's representer is $\alpha_j=\epsilon_ja_j$. Then
\[
 a_{1,0}=-cS_1,\qquad a_{2,0}=0,\quad
 \phi=q\{(Y-f_1)^2-(Y-f_2)^2\}+a_1(Y-f_1)-a_2(Y-f_2).
\]
The exact population contrast is
\[
 \theta_0=\tfrac12(b_2^2-b_1^2-2b_1c).
\]
We choose $b_2>0$ so that $\theta_0=\delta/\sqrt n$.
The five-refit baseline uses $\delta\in\{-0.7,-0.35,0,0.35,0.7\}$;
other schedules and the $p=256$ history comparison use $\delta\in\{-0.7,0,0.7\}$.
Thus equality of risk holds between distinct population forecasts, with
positive population score variance.

Forecasts minimize half mean squared training error plus an $\ell_1$
penalty. A second lasso regression fits $2q(\widehat f_j-Y)$ on the same
regressors to estimate $a_j$. Both representers are estimated, including
the zero population representer of the second rule. All coefficients,
including the intercept, are penalized.

For each forecast and correction, three chronological validation blocks
occupy fractions $[1/2,2/3)$, $[2/3,5/6)$ and $[5/6,1]$ of the eligible
history. Training prefixes end $b_n$ observations before their validation
blocks. Candidate penalties are
\[
 \lambda(\zeta)=\zeta\widehat\sigma_y
     \sqrt{2\log(p+1)/T'},\qquad
 \zeta\in\{1/32,1/16,1/8,1/4,1/2,1,2\},
\]
where $T'$ is the prefix length and $\widehat\sigma_y$ is the standard
deviation of the regression response, including the generated response
in the correction regression. The multiplier minimizes total validation
squared error, with ties favoring the larger penalty; the final fit
rescales it to the full training window. For correction validation, a
preliminary forecast fitted and tuned on the prefix generates the
responses in both training and validation samples. Its internal folds
use the same gap. The final correction uses responses generated by the
final forecast on the complete training window. Hence tuning uses no
future evaluation outcome. The performance of CV penalties does not
verify the learning-rate assumptions of Corollary~\external{cor:lasso}.

\subsection{Nonlinear probability forecasts and cost-weighted evaluation}
\label{app:sim-neural}

Let $Z_t\in\R^{10}$ be stationary with independent Gaussian AR(1)
coordinates, each with coefficient $\varrho$ and innovation variance
$1-\varrho^2$. Transform each coordinate by $2\Phi(Z_{it})-1$, where
$\Phi$ is the standard normal distribution function, and partition the
result into $U_t,V_t\in[-1,1]^5$. At a given date all ten coordinates are
independent and uniform. Define the centered functions
\begin{align*}
 Q(x)&=(3x^2-1)/2, & Q_4(x)&=(5x^4-1)/4,\\
 H(x,y)&=(9x^2y^2-1)/8, & R(x,y)&=3(x^2+y^2)/4-1/2,\\
 S(x,y)&=3(x+y)^2/10-1/5.
\end{align*}
The two signals are
\begin{align*}
 g_1(u)&=0.5Q(u_1)+0.3H(u_2,u_3)+0.2R(u_4,u_5),\\
 g_2(v)&=0.5Q_4(v_1)+0.3H(v_2,v_3)+0.2S(v_4,v_5).
\end{align*}
All five inputs enter each signal. Conditional on the predictor path,
independent binary outcomes have probabilities
\[
 \pi_t=\mu+a g_1(U_t)+b g_2(V_t),\qquad \mu=0.25,\quad a=0.24.
\]
Rule 1 observes $U_t$ and rule 2 observes $V_t$. Their population
conditional probabilities are $p_{1,0}=\mu+a g_1$ and
$p_{2,0}=\mu+b g_2$, since both signals have mean zero.
Training uses squared loss, whereas evaluation uses
$\ell_\tau(y,p)=(1+\tau y)(y-p)^2$ with $\tau=4$.
The analytic correction coefficient is
$A_\tau(p)=-2\tau p(1-p)$, as in main-paper
Example~\external{ex:probability-evaluation}. The corrected contrast evaluates
$\ell_\tau(Y,p)+A_\tau(p)(Y-p)$ at the two fitted probabilities.
Although no auxiliary regression is needed, estimating the probabilities
also introduces error into the correction coefficient.

\paragraph{Exact nulls and alternatives.}
Write $v_j=\E g_j^2$ and $t_j=\E g_j^3$, integrating independent uniform
coordinates. At a true conditional probability $p$, expected loss is
$(1+\tau)p-(1+2\tau)p^2+\tau p^3$. Therefore
\[
 \theta_0=K(b^2v_2-a^2v_1)+\tau(a^3t_1-b^3t_2),
 \qquad K=1+2\tau-3\tau\mu=6.
\]
We compute the polynomial moments by Gauss--Legendre quadrature, which is
exact here up to numerical precision, and choose $b>0$ to give
$\theta_0=\delta/\sqrt n$. Under the null, $b\simeq0.3188$.
The power grid is $\delta\in\{\pm0.1,\pm0.2,\pm0.4,\pm0.6\}$.
All configurations satisfy $0.02\le\pi_t\le0.95$. Opposite signs change
the signal differently: $\delta=-0.6$ weakens the second signal, whereas
$\delta=0.6$ brings the largest joint probability close to the upper
support restriction. This asymmetry is relevant to the power comparisons.

\paragraph{Training and regularization.}
Each forecasting rule has two hidden ReLU layers of width eight, an affine
skip connection and output map $0.5+0.49\tanh(z)$, for 134 parameters.
All parameters lie in $[-3,3]$. The ReLU activation is confined to the
hidden layers; the output map keeps fitted probabilities bounded.
The training objective is half mean squared error plus
$\lambda\sum_{i\ne i_0}\vartheta_i^2/2$, excluding only the affine
output intercept, with
\[
 \lambda=\kappa/\sqrt T,\qquad
 \kappa\in\{0.003,0.03,0.3,3\}.
\]
The penalties decrease with $T$; the population risk contrast remains
defined by the conditional probabilities $p_{1,0}$ and $p_{2,0}$.

At each refit, validation uses the last $\lceil T/4\rceil$ observations
of the eligible history, separated from the fitting prefix by $b_n^{\rm N}$.
Validation MSE selects the penalty and an iteration horizon in
$\{100,200,400,600\}$. Ties within $10^{-12}$ favor the larger penalty
and then the earlier horizon. Each candidate uses two initializations;
at each horizon, the fit with the lower training objective is evaluated
on the validation sample. The final refit uses all $T$ observations,
the selected penalty and horizon, and two new initializations.
Optimization uses PyTorch's Adam implementation with full-sample updates
and a cosine learning-rate schedule from 0.003 to 0.0003 over 600 steps.
The retained iterate minimizes the training objective among the initial
value, every tenth iterate and the relevant horizon. Training uses
single precision and inference double precision.

The chosen penalty $\kappa/\sqrt T$ differs from the
$\kappa/T$ sufficient calibration in Proposition~\ref{prop:nn-validation}.
The experiment therefore assesses finite-sample inference for this training
procedure without verifying that proposition's rate conditions or the
global optimization-gap assumption in Corollary~\external{cor:nn-rates}.
Its fixed architecture and positive penalty grid also restrict its scope:
it does not compare depths or unregularized networks.

\subsection{Additional size evidence}\label{app:sim-training}

\paragraph{Lasso: training information and refitting.}
The size comparison identifies training accuracy as an important limit
to correction. In Table~\ref{tab:expanded-lasso-size}, $T=2n$ brings
corrected rejection to 5.6--6.4\%, while raw rejection remains
20.9--31.8\%. With $T=0.5n$, correction
reduces rejection by 45--67 percentage points but leaves rates of
9.9--13.2\%. Bootstrapping raw scores retains substantial distortion.

\begin{table}[!htbp]
\centering\small
\begin{singlespace}
\setlength{\tabcolsep}{4pt}
\caption{Lasso size by training history and number of regressors}
\label{tab:expanded-lasso-size}
\medskip
\begin{tabular}{rrrrrr}
\toprule
$T/n$ & $p$ & Raw & Boot. Raw & Orth. & Boot. Orth. \\
\midrule
0.5 & 32 & 54.9 & 49.9 & 9.9 & 8.5 \\
0.5 & 128 & 74.8 & 68.7 & 12.9 & 11.1 \\
0.5 & 256 & 80.5 & 74.9 & 13.2 & 10.9 \\
1 & 32 & 33.9 & 30.8 & 7.2 & 7.3 \\
1 & 128 & 46.2 & 43.1 & 7.5 & 6.8 \\
1 & 256 & 53.6 & 48.7 & 8.6 & 8.0 \\
2 & 32 & 20.9 & 19.6 & 5.7 & 5.3 \\
2 & 128 & 27.7 & 26.4 & 5.6 & 5.3 \\
2 & 256 & 31.8 & 29.2 & 6.4 & 5.3 \\
\bottomrule
\end{tabular}
\par\medskip
\begin{minipage}{0.98\textwidth}
\footnotesize Rejection percentages at nominal 5\%; $n=1000$, five fits,
chronological CV and 1,000 replications per row. Boot. denotes the centered
score bootstrap with fitted forecasts and corrections held fixed.
All reported statistics are well defined.
\end{minipage}
\end{singlespace}
\end{table}

The gains persist across the refitting schedules at $n=500$ and
$n=1000$. At $n=500$, corrected rejection is 8.3--9.3\% under one fit,
five fits and refitting every 25 dates, against raw rejection of
35.3--39.6\%. Refitting at every date gives 5.2\% and 36.4\%, respectively,
based on 250 replications. The greater Monte Carlo uncertainty in this
comparison prevents a precise ranking of refitting schedules.

Table~\ref{tab:lasso-tuning} examines whether the correction's benefit
persists under an alternative penalty selector. On the same penalty
grid, this selector minimizes the working Gaussian criterion
$T\log(\mathrm{MSE})+2\operatorname{rank}(X_{\widehat S})$, where
$\widehat S$ is the active set, separately for forecasts and corrections;
ties favor the larger penalty. No information-criterion guarantee is
asserted for the dependent observations or generated responses.
Correction reduces rejection at $p=32$, but need not do so at $p=256$:
with $T=n$, Raw and Orthogonal reject 44.4\% and 47.1\%.
Thus the improvement under chronological CV does not extend to every
tuning rule; orthogonality still requires sufficiently accurate estimates.

\begin{table}[!htbp]
\centering\footnotesize
\begin{singlespace}
\setlength{\tabcolsep}{3pt}
\caption{Lasso size under an alternative training-only selector}
\label{tab:lasso-tuning}
\medskip
\begin{tabular}{rrlrrrr}
\toprule
$T/n$ & $p$ & Tuning & Raw & Boot. Raw & Orth. & Boot. Orth. \\
\midrule
0.5 & 32 & CV & 54.9 & 49.9 & 9.9 & 8.5 \\
0.5 & 32 & AIC & 17.7 & 13.8 & 12.3 & 11.6 \\
0.5 & 256 & CV & 80.5 & 74.9 & 13.2 & 10.9 \\
0.5 & 256 & AIC & 72.7 & 67.0 & 73.1 & 66.8 \\
1 & 32 & CV & 33.9 & 30.8 & 7.2 & 7.3 \\
1 & 32 & AIC & 13.7 & 12.3 & 8.0 & 7.4 \\
1 & 256 & CV & 53.6 & 48.7 & 8.6 & 8.0 \\
1 & 256 & AIC & 44.4 & 38.5 & 47.1 & 42.1 \\
2 & 32 & CV & 20.9 & 19.6 & 5.7 & 5.3 \\
2 & 32 & AIC & 9.9 & 9.8 & 6.2 & 5.9 \\
2 & 256 & CV & 31.8 & 29.2 & 6.4 & 5.3 \\
2 & 256 & AIC & 19.1 & 17.8 & 20.2 & 18.7 \\
\bottomrule
\end{tabular}
\par\medskip
\begin{minipage}{0.98\textwidth}
\footnotesize Rejection percentages at nominal 5\%; $n=1000$, five fits
and 1,000 replications per row. CV denotes chronological validation;
AIC denotes the stated working information criterion on the eligible
training sample. The population forecasts and targets are unchanged.
All reported statistics are well defined.
\end{minipage}
\end{singlespace}
\end{table}

\paragraph{Neural forecasts: covariance and bootstrap sensitivity.}
Correction reduces neural size distortion most clearly at
$n=1000,T=4000$ (Table~\ref{tab:nn-size}). Raw and bootstrap raw rejection
are both 14.7\%, compared with 7.8\% and 7.9\% for the corrected Wald
and bootstrap tests. At $n=500,T=4000$, the reduction from 11.3\% to
8.8\% is smaller, and the comparison with Bootstrap Raw is inconclusive.
At $n=1000,T=8000$, the reduction from 8.8\% to 7.3\% is also less
precisely estimated. For Orthogonal versus Raw, the paired McNemar $p$-values are 0.006
at $(n,T)=(500,4000)$, less than $10^{-12}$ at $(1000,4000)$, and
0.082 at $(1000,8000)$, without adjustment for multiple comparisons. All three corrected Wald
frequencies remain significantly above 5\% in pointwise comparisons:
correction reduces distortion but does not restore nominal size over
this grid.

\begin{table}[!htbp]
\centering\small
\begin{singlespace}
\setlength{\tabcolsep}{4pt}
\caption{Size of the four probability-forecast tests}
\label{tab:nn-size}
\medskip
\begin{tabular}{rr*{4}{r}}
\toprule
$n$ & $T$ & Raw & Boot. Raw & Orth. & Boot. Orth.\\
\midrule
500 & 4,000 & 11.3 & 10.1 & 8.8 & 8.5\\
1,000 & 4,000 & 14.7 & 14.7 & 7.8 & 7.9\\
1,000 & 8,000 & 8.8 & 9.4 & 7.3 & 6.6\\
\bottomrule
\end{tabular}
\par\medskip
\begin{minipage}{0.98\textwidth}
\footnotesize Rejection percentages at nominal 5\%; two hidden layers of width eight,
five fits and 1,024 replications per row. Boot. denotes the score bootstrap
with 999 resamples and primary block length $2(L_n+1)$. All methods
estimate covariance and all reported statistics are well defined.
\end{minipage}
\end{singlespace}
\end{table}

The central size advantage persists under the covariance and resampling
choices in Table~\ref{tab:nn-sensitivity}. The covariance checks replace
$L_n$ by $\lfloor L_n/2\rfloor$ or $2L_n$; the bootstrap checks retain
$L_n$ and use block lengths $L_n+1$ or $4(L_n+1)$ instead of
$2(L_n+1)$. Longer blocks increase corrected rejection to 11.3\%, 9.5\%
and 8.0\% in the three null settings. Bootstrap calibration can therefore
worsen the remaining distortion. These bandwidths and block lengths
were fixed before observing the corresponding rejection frequencies.

\begin{table}[!htbp]
\centering\small
\begin{singlespace}
\setlength{\tabcolsep}{4pt}
\caption{Neural size under covariance-bandwidth and bootstrap-block changes}
\label{tab:nn-sensitivity}
\medskip
\begin{tabular}{l*{6}{r}}
\toprule
$(n,T)$ & \multicolumn{2}{c}{$(500,4{,}000)$} & \multicolumn{2}{c}{$(1{,}000,4{,}000)$} & \multicolumn{2}{c}{$(1{,}000,8{,}000)$}\\
Calibration & Raw & Orth. & Raw & Orth. & Raw & Orth.\\
\midrule
HAC half bandwidth & 10.4 & 7.8 & 14.8 & 7.9 & 8.3 & 6.8\\
HAC baseline & 11.3 & 8.8 & 14.7 & 7.8 & 8.8 & 7.3\\
HAC double bandwidth & 11.2 & 9.8 & 15.7 & 8.4 & 9.8 & 7.9\\
Bootstrap short blocks & 9.9 & 8.0 & 14.1 & 7.6 & 8.6 & 6.8\\
Bootstrap baseline blocks & 10.1 & 8.5 & 14.7 & 7.9 & 9.4 & 6.6\\
Bootstrap long blocks & 12.8 & 11.3 & 15.6 & 9.5 & 10.4 & 8.0\\
\bottomrule
\end{tabular}
\par\medskip
\begin{minipage}{0.98\textwidth}
\footnotesize Rejection percentages with 1,024 null replications per setting.
Baseline HAC bandwidths are 18 at $n=500$ and 20 at $n=1000$.
Half and double bandwidths are $\lfloor L_n/2\rfloor$ and $2L_n$.
Short, baseline and long bootstrap blocks have lengths $L_n+1$,
$2(L_n+1)$ and $4(L_n+1)$; all bootstrap rows use the baseline
HAC bandwidth and 999 resamples. All choices were fixed before the
extension; no rejection-based selection is used.
\end{minipage}
\end{singlespace}
\end{table}

\subsection{Local power and empirical size adjustment}\label{app:sim-power}

To distinguish power from size distortion, we report rejection frequencies
using both the nominal critical values and critical values calibrated to
each method's matching empirical null distribution. For method $m$, let
$A_{m,r}$ be the absolute studentized statistic for a Wald test and the
negative bootstrap $p$-value for a bootstrap test. From $R_0$ null replications, choose the empirical
95th percentile $c_m$ and a boundary weight $\omega_m\in[0,1]$ so that
\[
 \frac1{R_0}\sum_{r=1}^{R_0}
 \{\ind(A_{m,r}^0>c_m)+\omega_m\ind(A_{m,r}^0=c_m)\}=0.05.
\]
Size-adjusted power averages the same expression over the alternative
replications. The boundary weight accommodates discrete bootstrap
$p$-values and noninteger 5\% tail counts. This adjustment uses the simulated null distribution and is a diagnostic,
rather than an additional feasible testing procedure. Its empirical null
rejection frequency is 5\% by construction; alternative draws play no role
in selecting the critical value.

Monte Carlo uncertainty accounts for estimation of the null critical values
and for shared randomness across methods and null/alternative configurations.
We jointly resample shared replication indices, separately resample the
additional null indices, and reestimate each critical value in 1,000
resamples. This includes uncertainty omitted by binomial standard errors
conditional on a fixed cutoff. Intervals and paired tests are pointwise,
without adjustment for multiple comparisons.

\begin{figure}[!htbp]
\centering
\includegraphics[width=\textwidth]{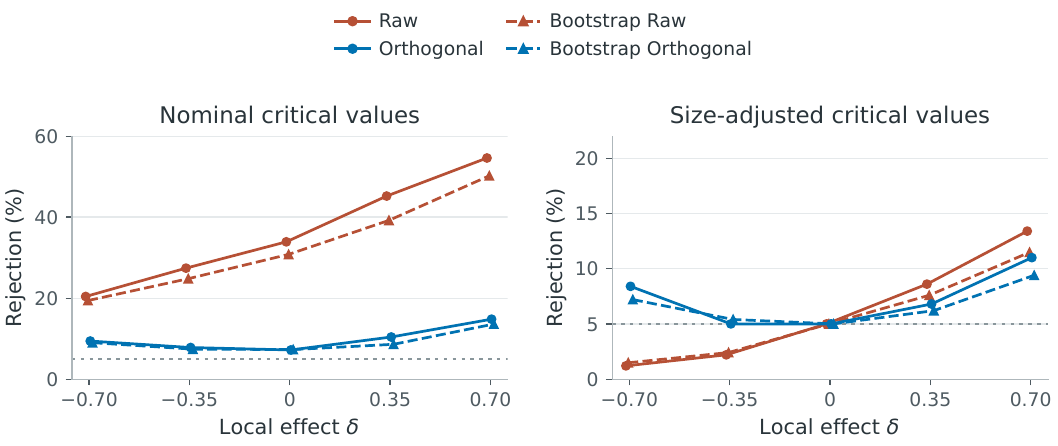}
\caption{Lasso local power with Wald and bootstrap tests.}
\label{fig:lasso-power-four}
\begin{singlespace}\small\raggedright
\noindent Notes: Chronological CV, $n=T=1000$, $p=32$, five fits,
1,000 null and 500 replications per nonzero alternative. The target is
$\delta/\sqrt n$. Size adjustment uses each method's own empirical null
distribution, including the bootstrap $p$-value distribution. Adjusted
null points equal 5\% by construction. Curves connect simulated settings.
\end{singlespace}
\end{figure}

\paragraph{Lasso.}
Positive estimation drift makes the raw test's sensitivity depend strongly
on the sign of the alternative. Figure~\ref{fig:lasso-power-four} illustrates
this at $n=T=1000,p=32$ with five fits. At $\delta=-0.7$, the alternative
offsets the drift and lowers raw rejection from 33.9\% under the null to
20.4\%; at $\delta=0.7$, raw rejection rises to 54.6\%.
The corrected frequencies are 9.4\% and 14.8\%. After size adjustment,
Raw has power 1.2\% and 13.4\%, compared with 8.4\% and 11.0\% for
Orthogonal. Correction restores sensitivity in the negative direction,
although power remains modest and the gain is not uniform across signs.
The paired adjusted-power gain over Raw at $\delta=-0.7$ is 7.2
percentage points, with a 95\% Monte Carlo interval of $[3.8,9.6]$;
at $\delta=0.7$, the difference is $-2.4$ points with interval
$[-6.6,1.4]$.

\begin{figure}[!htbp]
\centering
\includegraphics[width=\textwidth]{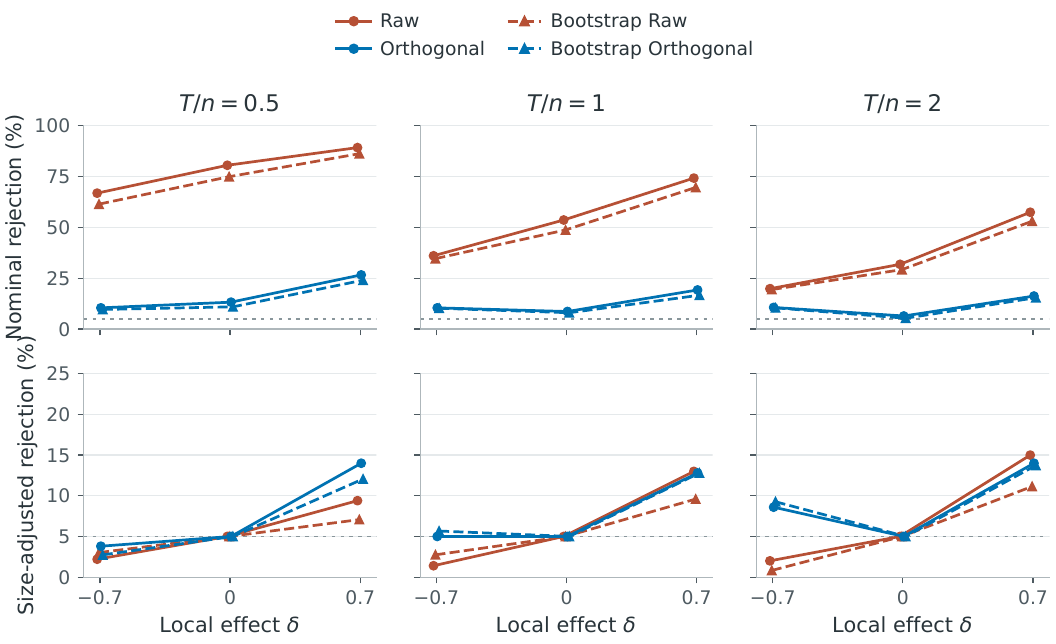}
\caption{Lasso local power across training histories with more regressors.}
\label{fig:expanded-complexity-power}
\begin{singlespace}\small\raggedright
\noindent Notes: Chronological CV, $n=1000$, $p=256$, five fits,
1,000 null and 500 replications per nonzero alternative. The four tests
and null-calibration rule are the same as in Figure~\ref{fig:lasso-power-four}.
\end{singlespace}
\end{figure}

Limited training also affects sensitivity after size adjustment.
Figure~\ref{fig:expanded-complexity-power} considers $p=256$ across
all three training ratios. At $T=0.5n$, corrected adjusted power is
3.8\% at $\delta=-0.7$ and 14.0\% at $\delta=0.7$.
With $T=2n$, these frequencies are 8.6\% and 14.0\%, compared with
raw adjusted power of 2.0\% and 15.0\%. The shorter history thus leaves
a pronounced directional imbalance. The same pattern appears at the smaller evaluation size and under the
three principal refitting schedules. The replication files report these
additional four-method comparisons.

\begin{figure}[!htbp]
\centering
\includegraphics[width=\textwidth]{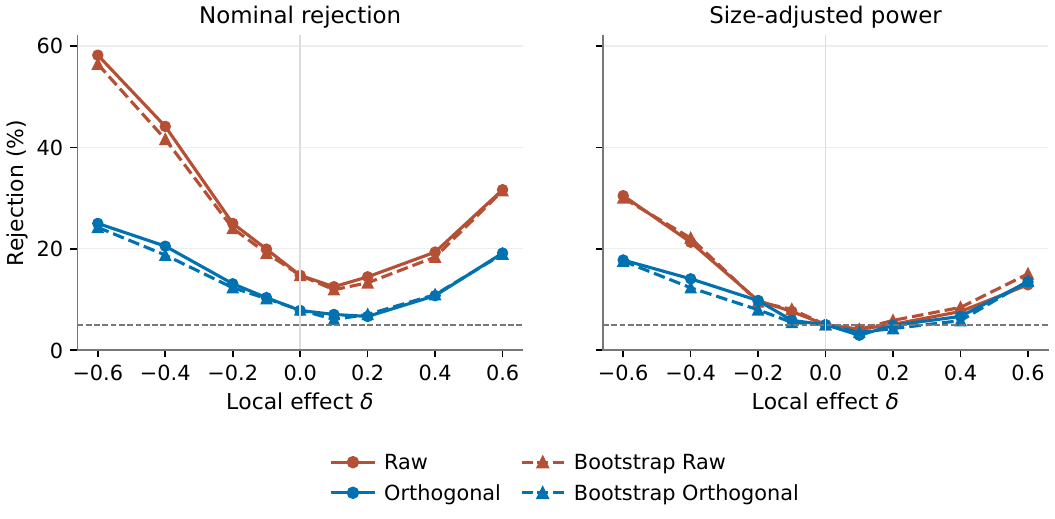}
\caption{Nonlinear probability forecasts: nominal and size-adjusted local power.}
\label{fig:nn-power}
\begin{singlespace}\small\raggedright
\noindent Notes: $n=1000$, $T=4000$, two hidden layers of width eight,
five fits and training-only selection of penalties and horizons. Each
alternative has 512 replications and the common matched null 1,024.
Adjusted power uses method-specific cutoffs; its null point equals 5\%
by construction. Positive and negative alternatives change signal strength
differently, as described in Section~\ref{app:sim-neural}.
\end{singlespace}
\end{figure}

\paragraph{Neural probabilities.}
The neural size improvement entails a power loss at the stronger negative
alternatives (Figure~\ref{fig:nn-power}). At $\delta=-0.6$, size-adjusted
power is 30.5\% for Raw, 30.0\% for Bootstrap Raw, 17.8\% for Orthogonal
and 17.6\% for Bootstrap Orthogonal. At $\delta=-0.4$, the corresponding
frequencies are 21.3\%, 22.1\%, 14.1\% and 12.3\%.
The paired intervals in Table~\ref{tab:nn-power-comparisons} exclude zero
for Orthogonal versus both raw-score tests at these two alternatives. At $\delta=0.6$, adjusted raw
and corrected power are 12.9\% and 13.5\%; method differences at the
other alternatives are generally less precisely estimated.
The figure includes all eight alternatives, preserving the asymmetry
between their positive and negative signs.

\begin{table}[!htbp]
\centering\small
\begin{singlespace}
\setlength{\tabcolsep}{4pt}
\caption{Paired differences in size-adjusted neural power}
\label{tab:nn-power-comparisons}
\medskip
\begin{tabular}{r*{2}{r}}
\toprule
$\delta$ & Orth. $-$ Raw & Orth. $-$ Boot. Raw\\
\midrule
-0.6 & -12.7 [-18.4, -7.2] & -12.2 [-16.6, -6.0]\\
-0.4 & -7.2 [-11.9, -3.1] & -8.0 [-11.6, -3.3]\\
-0.2 & 0.0 [-3.9, 2.3] & 0.3 [-2.6, 3.2]\\
-0.1 & -1.8 [-4.1, 0.4] & -2.2 [-3.9, 0.7]\\
+0.1 & -1.2 [-2.7, 1.0] & -1.2 [-2.5, 1.1]\\
+0.2 & -0.2 [-2.1, 1.6] & -1.0 [-2.7, 1.2]\\
+0.4 & -1.0 [-3.7, 1.6] & -1.8 [-4.0, 1.1]\\
+0.6 & 0.6 [-2.5, 4.1] & -1.6 [-5.1, 3.3]\\
\bottomrule
\end{tabular}
\par\medskip
\begin{minipage}{0.98\textwidth}
\footnotesize Differences in percentage points, with pointwise 95\% percentile
intervals in brackets. Settings have $n=1000,T=4000$, 512 alternative
and 1,024 null replications. The 1,000 resamples preserve pairing of
methods and shared null/alternative paths, stratify the additional null
paths, and reestimate each method's null cutoff. Intervals are not
adjusted for multiple comparisons.
\end{minipage}
\end{singlespace}
\end{table}

\subsection{Learning error, population drift and studentization}
\label{app:sim-diagnostics}

The diagnostics distinguish changes in population mean error from changes
in sampling variation. For fixed fitted objects in window $k$, let
$d_{a,k}$ be the raw or corrected contrast and define
\[
 D_a=\sqrt n\sum_k w_k\{P d_{a,k}-\theta_0\},\qquad
 w_k=|I_k|/n,\qquad a\in\{\mathrm{raw},\mathrm{orth}\}.
\]
Here $P$ integrates over the marginal population distribution with fitted
objects fixed, rather than the distribution of evaluation observations
conditional on training. This distinction matters under temporal dependence.
A small mean signed $D_a$ can also reflect cancellation, so it does not
verify the simultaneous learning conditions.

\paragraph{Lasso.}
Here $a_{j,0}$ and $\widehat a_{jk}$ denote the coefficient vectors
of the population and fitted correction functions. Write
$u_{jk}=\widehat\beta_{jk}-\beta_{j,0}$ and
$v_{jk}=\widehat a_{jk}-a_{j,0}$, and put
$Q(u)=\tfrac12\norm{u}_2^2+u_0u_W$, with $0,W$ identifying the
intercept and state coordinates. Exact population integration gives
\begin{align*}
 D_{\rm raw}
 &=\sqrt n\sum_k w_k\sum_j\epsilon_j
       \{a_{j,0}^{\T}u_{jk}+Q(u_{jk})\},\\
 D_{\rm orth}
 &=\sqrt n\sum_k w_k\sum_j\epsilon_j
       \{Q(u_{jk})-v_{jk}^{\T}u_{jk}\}.
\end{align*}
Correction removes the linear forecast-error term, leaving quadratic
forecast error and a product of forecast and representer errors.
Population $L^2$ learning errors equal coefficient Euclidean errors
because the Gram matrix is the identity. Recorded maxima are taken over
both rules and all refitting windows within a replication.

\begin{figure}[!htbp]
\centering
\includegraphics[width=\textwidth]{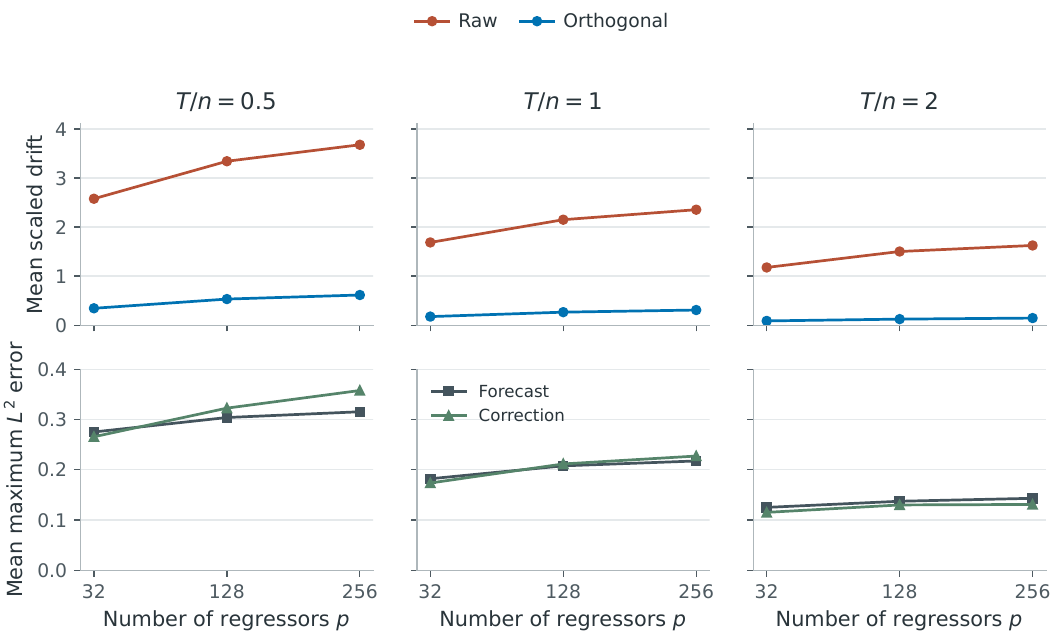}
\caption{Lasso population drift and learning error.}
\label{fig:expanded-lasso-diagnostics}
\begin{singlespace}\small\raggedright
\noindent Notes: Chronological CV, $n=1000$, five fits and 1,000
replications per configuration. Drift uses exact integration over the marginal population distribution,
with fitted objects fixed. Learning errors are mean maximum
population $L^2$ errors of the forecasts and representers, with maxima
taken over both rules and all training windows within each replication.
\end{singlespace}
\end{figure}

The lasso size improvement is accompanied by a substantial reduction
in drift (Figure~\ref{fig:expanded-lasso-diagnostics} and
Table~\ref{tab:expanded-learning}). At $p=32,T=n$, mean scaled drift
falls from 1.687 to 0.174 and 95\% coverage rises from 66.1\% to 92.8\%,
while mean interval length changes little, from 0.139 to 0.138.
At $p=256,T=0.5n$, corrected drift remains 0.615. The error expansion
shows that both squared forecast error and the forecast--representer
error product can contribute to the remaining distortion.

\begin{table}[!htbp]
\centering\small
\begin{singlespace}
\setlength{\tabcolsep}{4pt}
\caption{Lasso learning errors, population drift and confidence coverage}
\label{tab:expanded-learning}
\medskip
\begin{tabular}{rrrrrrrr}
\toprule
$T/n$ & $p$ & Forecast & Repr. & $D_{\rm raw}$ & $D_{\rm orth}$ & Raw cov. & Orth. cov. \\
\midrule
0.5 & 32 & 0.275 & 0.266 & 2.578 & 0.344 & 45.1 & 90.1 \\
0.5 & 128 & 0.304 & 0.323 & 3.343 & 0.532 & 25.2 & 87.1 \\
0.5 & 256 & 0.316 & 0.358 & 3.679 & 0.615 & 19.5 & 86.8 \\
1 & 32 & 0.182 & 0.174 & 1.687 & 0.174 & 66.1 & 92.8 \\
1 & 128 & 0.208 & 0.212 & 2.151 & 0.265 & 53.8 & 92.5 \\
1 & 256 & 0.218 & 0.228 & 2.355 & 0.308 & 46.4 & 91.4 \\
2 & 32 & 0.125 & 0.115 & 1.177 & 0.086 & 79.1 & 94.3 \\
2 & 128 & 0.138 & 0.130 & 1.502 & 0.123 & 72.3 & 94.4 \\
2 & 256 & 0.143 & 0.131 & 1.625 & 0.144 & 68.2 & 93.6 \\
\bottomrule
\end{tabular}
\par\medskip
\begin{minipage}{0.98\textwidth}
\footnotesize Mean maximum $L^2$ forecast and representer errors, mean scaled population drift, and coverage percentages of nominal 95\% HAC intervals; $n=1000$, five refits and 1,000 replications. Coverage targets the population risk contrast. Error maxima are taken over both rules and all windows within each replication.
\end{minipage}
\end{singlespace}
\end{table}

\paragraph{Neural probabilities.}
For one rule, put $p=p_{j,0}$ and $e=\widehat p_j-p$. Conditional mean
integration gives the exact per-rule mean discrepancies
\[
 P\{A_\tau(p)e+(1+\tau p)e^2\},\qquad
 P\{(1+2\tau-3\tau p)e^2-2\tau e^3\}
\]
for the raw and corrected scores, respectively. Contrasting the two rules,
averaging over windows and multiplying by $\sqrt n$ gives
$D_{\rm raw}$ and $D_{\rm orth}$.
The linear term vanishes after correction, but nonlinear forecast error
remains. These integrals are evaluated on 4,096 independent uniform
predictor draws per replication, integrating binary outcomes analytically.
The diagnostic reports the mean absolute $D_a$ across replications
within each null configuration. This differs from the absolute value
of the mean signed drift.

Both mean error and studentization contribute to the neural results.
At $n=1000,T=4000$ under the null, mean absolute scaled drift falls
from 0.160 to 0.113.
Table~\ref{tab:nn-learning} shows that longer training reduces forecast
and correction errors; on average, both rules also improve on their
constant validation benchmark. The raw test's root mean squared estimated
standard error is 0.00921, against a Monte Carlo standard deviation of
0.01201; the corresponding corrected values are 0.01167 and 0.01268.
Reported uncertainty is therefore closer to sampling dispersion after
correction, although signed average bias does not uniformly decrease.
These diagnostics suggest several contributors to the improvement
rather than a unique explanation of the remaining distortion.

\begin{figure}[!htbp]
\centering
\includegraphics[width=\textwidth]{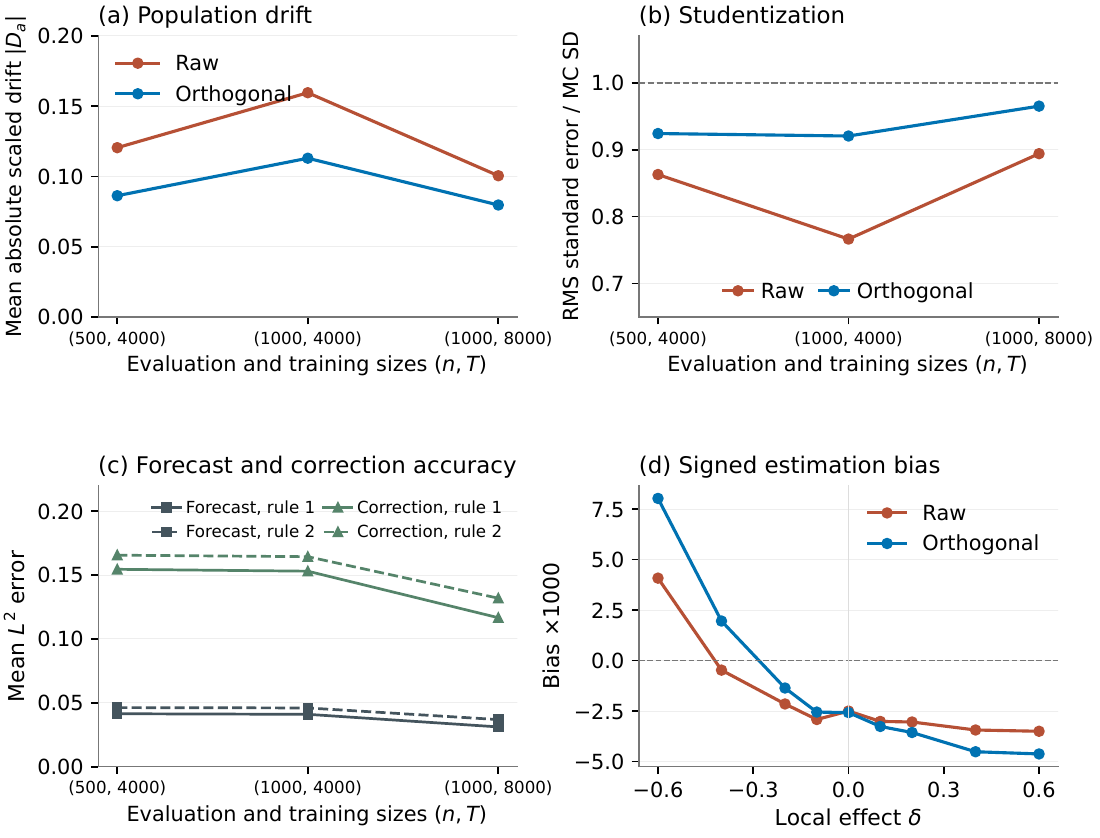}
\caption{Neural learning error, uncertainty estimates and attenuation of local alternatives.}
\label{fig:nn-diagnostics}
\begin{singlespace}\small\raggedright
\noindent Notes: The null panels use 1,024 replications per setting and
show mean absolute population drift $D_a$, root mean squared estimated
standard error divided by empirical standard deviation, and mean population
$L^2$ forecast and correction errors for both rules. The bias panel uses
512 replications per alternative at $n=1000,T=4000$. Population diagnostics
use independent predictor draws with fitted objects held fixed.
\end{singlespace}
\end{figure}

\begin{table}[!htbp]
\centering\small
\begin{singlespace}
\setlength{\tabcolsep}{4pt}
\caption{Learning accuracy for the nonlinear probability forecasts}
\label{tab:nn-learning}
\medskip
\begin{tabular}{rrr*{4}{r}}
\toprule
$n$ & $T$ & Rule & Rel. RMSE & Forecast & Correction & Val. gain\\
\midrule
500 & 4,000 & 1 & 0.708 & 0.041 & 0.155 & 2.084\\
500 & 4,000 & 2 & 0.786 & 0.046 & 0.166 & 1.837\\
1,000 & 4,000 & 1 & 0.703 & 0.041 & 0.153 & 2.143\\
1,000 & 4,000 & 2 & 0.787 & 0.046 & 0.164 & 1.813\\
1,000 & 8,000 & 1 & 0.533 & 0.031 & 0.117 & 2.528\\
1,000 & 8,000 & 2 & 0.626 & 0.037 & 0.132 & 2.242\\
\bottomrule
\end{tabular}
\par\medskip
\begin{minipage}{0.98\textwidth}
\footnotesize Null configurations, 1,024 replications and five fits per rule.
Relative RMSE is the median across replications of mean-window forecast
RMSE divided by the population signal standard deviation. Forecast and
correction columns average the population $L^2$ errors of $\widehat p_j$
and $A_\tau(\widehat p_j)$ over replications and windows. Population errors
use 4,096 independent predictor draws. Validation gain is 1,000 times the difference between constant-forecast
and fitted-forecast validation MSE; positive values favor the fitted
forecast.
\end{minipage}
\end{singlespace}
\end{table}

The bias diagnostics help interpret the power loss. At $\delta=-0.6$,
the corrected estimator's bias is about 0.00803 against a true contrast
of $-0.01897$, shifting estimates toward the null. This attenuation is
consistent with the remaining nonlinear forecast-error terms in the
mean expansion. The diagnostics do not separately attribute forecast
error to approximation, regularization or optimization, and do not verify
the global objective-gap or asymptotic inference conditions.

\subsection{Score and covariance replacement diagnostics}
\label{app:sim-replacement}

The population drift above is only one component of the difference between
feasible and oracle inference. Let
$\widehat\theta_n^{\mathrm{or}}=n^{-1}\sum_t\phi_t(\nu_0)$ use the
population forecasts and representers on the same evaluation observations,
and define
\[
 O_n=\sqrt n(\widehat\theta_n^{\mathrm{or}}-\theta_0),\qquad
 C_n=\sqrt n(\widehat\theta_n-\widehat\theta_n^{\mathrm{or}})
          -D_{\mathrm{orth}}.
\]
Then the exact decomposition is
\begin{equation}\label{eq:simulation-replacement}
 \sqrt n(\widehat\theta_n-\theta_0)=O_n+D_{\mathrm{orth}}+C_n.
\end{equation}
Here $D_{\mathrm{orth}}$ integrates over the marginal population law
with fitted objects fixed, as in Section~\ref{app:sim-diagnostics}.
Thus $C_n$ is the centered score-replacement remainder in
\eqref{eq:centeredreplacement}. It measures the remaining discrepancy
from using fitted scores after subtracting their population drift; it
need not have zero finite-sample expectation under chronological fitting.

Table~\ref{tab:score-replacement} reports root mean squared (RMS)
drift and centered remainder, each divided by the Monte Carlo standard
deviation $s_O$ of $O_n$. To examine covariance replacement, compute
the feasible and oracle Bartlett estimates with the same bandwidth and put
$v_n=(\widehat\Omega_n-\widehat\Omega_n^{\mathrm{or}})/
\widehat\Omega_n^{\mathrm{or}}$. The table reports the Monte Carlo
averages of $v_n$ and $|v_n|$ in percent. All quantities use paired
saved replications; no forecasts are refitted for these diagnostics.
Lasso drift is integrated exactly. Neural drift is estimated using the
4,096 independent predictor draws described above, so its reported
centered remainder also contains numerical integration error.

\begin{table}[!htbp]
\centering\small
\begin{singlespace}
\setlength{\tabcolsep}{5pt}
\caption{Population drift, centered score replacement and covariance replacement}
\label{tab:score-replacement}
\medskip
\begin{tabular}{rrrrrrr}
\toprule
$n$ & $T$ & $K_n$ & $\mathrm{RMS}(D)/s_O$ & $\mathrm{RMS}(C)/s_O$ & $\overline v$ (\%) & $\overline{|v|}$ (\%) \\
\midrule
\multicolumn{7}{l}{Lasso: CV, $p=32$} \\
1,000 & 500 & 5 & 0.463 & 0.486 & -4.6 & 17.7 \\
1,000 & 1,000 & 5 & 0.260 & 0.338 & -8.4 & 14.2 \\
1,000 & 2,000 & 5 & 0.140 & 0.239 & -8.1 & 11.8 \\
1,000 & 1,000 & 1 & 0.347 & 0.340 & -8.0 & 17.2 \\
1,000 & 1,000 & 40 & 0.253 & 0.333 & -8.6 & 14.1 \\
\midrule
\multicolumn{7}{l}{Neural probabilities} \\
500 & 4,000 & 5 & 0.257 & 0.719 & -20.3 & 30.7 \\
1,000 & 4,000 & 5 & 0.344 & 0.763 & -20.8 & 28.1 \\
1,000 & 8,000 & 5 & 0.237 & 0.590 & -18.2 & 22.1 \\
\bottomrule
\end{tabular}
\par\medskip
\begin{minipage}{0.98\textwidth}
\footnotesize Null configurations. Lasso uses $p=32$, chronological CV
and 1,000 replications per row; neural forecasts use five fits and
1,024 replications per row. The $D$ and $C$ columns are
$\operatorname{RMS}(D_{\mathrm{orth}})/s_O$ and
$\operatorname{RMS}(C_n)/s_O$, where $s_O$ is the Monte Carlo standard
deviation of the corrected oracle sampling term. The $\bar v$ and
$\overline{|v|}$ columns average per-replication relative variance
discrepancies in percent; they are not ratios of average variances.
Forty fits correspond to refitting every 25 evaluation dates.
Neural drift and centered remainders include population-integration error.
\end{minipage}
\end{singlespace}
\end{table}

The lasso drift reduction does not eliminate the remaining score
discrepancy. At $n=T=1000$, $p=32$ and five fits, RMS raw drift is
$1.658s_O$, compared with $0.260s_O$ after correction. Holding $n=1000$
and five fits fixed, increasing $T$ from 500 to 2000 reduces normalized
corrected drift from 0.463 to 0.140 and the centered remainder from
0.486 to 0.239. Mean absolute relative covariance discrepancy falls
from 17.7\% to 11.8\%. At $T=1000$, the centered remainder is similar
across one, five and forty fits: 0.340, 0.338 and 0.333, respectively.
Longer training reduces these discrepancies, and the centered remainder
is similar across the refitting schedules considered.

The neural centered remainder is larger relative to oracle sampling
variation. At $n=1000$, increasing $T$ from 4000 to 8000 reduces its
normalized RMS magnitude from 0.763 to 0.590 and mean absolute relative
covariance discrepancy from 28.1\% to 22.1\%. Substantial discrepancies
therefore remain even after removing population drift.

These RMS ratios are not variance shares: the components of
\eqref{eq:simulation-replacement} are correlated. A nonzero $C_n$ also
arises with a single fit, so it does not isolate an effect of repeated
fitting or temporal dependence. Likewise, comparison with the oracle
Bartlett estimate measures replacement error, not the oracle estimate's
accuracy for the population long-run variance. The diagnostics support
drift reduction and the importance of learning accuracy at these sample
sizes; they do not verify asymptotic rates or uniquely attribute the
remaining size distortion.

\subsection{Scope and simulation design}\label{app:sim-support}
\label{app:sim-replay}

The experiments support size improvements for CV-tuned lasso across the
sample sizes, numbers of regressors and refitting schedules considered,
and for the nonlinear probability design. Their scope is limited by the remaining estimation
error, the neural power losses and the training procedures considered.
Both designs fix $\varrho=0.5$, and neither explores arbitrary dependence
strength or tuning choices; the lasso comparison also excludes $p>T$.

An analytic-penalty experiment illustrates a further limit to finite-sample
inference. Conservative calibration based on the known dependence envelope
and coefficient bounds can select identical zero forecasts and yield
degenerate variance. At $n=1000,T=0.5n$, only 46 of 1,000 corrected
statistics are well defined for $p=32$, and only two and one for $p=128$
and $p=256$. Rejection frequencies conditional on so few nondegenerate
statistics are uninformative. With $p=32,T=10n$, corrected rejection is
25.0\%, falling to 7.5\% at $T=50n$. This limitation concerns the particular
calibration used here, rather than analytic penalties in general.
The accompanying lasso results contain the full tuning comparisons.

Pilot simulations informed the neural architecture and training settings;
pilot draws are excluded from the reported results. Inspection of the
initial 2,560 production draws preceded an extension of 4,608 draws that
increased precision and added the stronger alternatives. The analysis
includes every initial production draw and every configuration in the
extension. It should therefore be interpreted as a study developed in
stages, not an entirely prespecified independent confirmation. Source
snapshots, seeds and individual draws document this sequence. No
configuration or replication is excluded because of its rejection
frequency or optimization diagnostics.

Pointwise Wilson intervals describe the precision of size estimates;
inference on method differences uses the paired draws, rather than the
overlap of marginal intervals. Power comparisons retain both signs and
account for estimated null critical values as in Section~\ref{app:sim-power}.
The reproducible analysis provides exact plotted values and four-method
summaries. Earlier neural, forest and kernel-ridge studies are archived separately and
supply no observations or claims to the evidence presented here.
\FloatBarrier
\begin{singlespace}
\bibliographystyle{plainnat}
\bibliography{references}
\end{singlespace}